\documentclass[12pt,journal,onecolumn,draftclsnofoot,twoside]{IEEEtran}
\usepackage{amsmath,amssymb,threeparttable,placeins,algorithm,algorithmic,multirow}
\usepackage{lineno}
\usepackage[dvips]{graphicx}
\usepackage[usenames]{color}
\usepackage{amsfonts}
\usepackage{latexsym}
\usepackage{subfigure}
\usepackage[justification=centering]{caption}

\newtheorem{theorem}{{{\textit{Theorem}}}}

\newtheorem{definition}{{{\textit{Definition}}}}
\newtheorem{proposition}{{{\textit{Proposition}}}}
\newtheorem{remark}{{{\textit{Remark}}}}

\def\BibTeX{{\rm B\kern-.05em{\sc i\kern-.025em b}\kern-.08em
    T\kern-.1667em\lower.7ex\hbox{E}\kern-.125emX}}
\begin{document}
	
	\title{Quasi-Orthogonal Z-Complementary Pairs and Their Applications in Fully Polarimetric Radar Systems}
	\author{Jiahuan~Wang,~
		Pingzhi~Fan,~\IEEEmembership{Fellow,~IEEE},
		Zhengchun~Zhou,~\IEEEmembership{Member,~IEEE},
		Yang~Yang~\IEEEmembership{Member,~IEEE}
		
	}

\maketitle
\begin{abstract}
 One objective of this paper is to propose a novel class of sequence pairs, called ``quasi-orthogonal Z-complementary pairs (QOZCPs)", each depicting Z-complementary property for their aperiodic auto-correlation sums and also having a low correlation zone when their aperiodic cross-correlation is considered. Construction of QOZCPs based on Successively Distributed Algorithms under Majorization Minimization (SDAMM) is presented.
 Another objective of this paper is to apply the proposed QOZCPs in fully polarimetric radar systems and analyse the corresponding ambiguity functions. It turns out that QOZCP waveforms are much more Doppler resilient than the known Golay complementary waveforms.

\end{abstract}


\section{Introduction}

This paper introduces a novel class of sequence pairs called
``quasi-orthogonal Z-complementary pairs (QOZCPs)" and their applications in designing Doppler resilient waveforms in polarimetric radar systems. In what follows, we first review the state-of-the-art
on pairs of sequences and Doppler resilient waveforms in polarimetric radar systems, then introduce our contributions
in this work.

\subsection{Sequence pairs}

Research on designing sequence pairs with good correlation properties started in early 1950's when M. J. Golay proposed Golay complementary pairs (GCPs), in his work on multislit spectrometry \cite{golay1951static}. GCPs are sequence pairs, each having zero aperiodic auto-correlation sums at every out-of-phase time-shifts \cite{golay1951static}. Since then a lot of works have been done on analysing the properties and systematic constructions of the GCPs \cite{golay1961complementary,Borwein2003,Davis1999,Feng1999}. However, binary GCPs are available only for lengths of the form $2^\alpha 10^\beta 26^\gamma$ (where $\alpha$, $\beta$ and $\gamma$ are non-negative integers) \cite{Borwein2003}. In search of binary sequence pairs of other lengths, depicting similar properties to that of the GCPs, Fan \textit{et al.} \cite{fan2007z} proposed binary Z-complementary pairs (ZCPs). ZCPs are sequence pairs having zero auto-correlation sums for each time-shifts within a certain region around the in-phase position, called the zero-correlation-zone (ZCZ) \cite{fan2007z}. ZCPs are available for many more lengths as compared to GCPs \cite{fan2007z}. Systematic constructions of ZCPs based on insertion method and generalised Boolean functions for even and odd-lengths have also been extensively studied \cite{Liu2014,Adhikary2020a,Adhikary2020,Chen2017,Shen2019,Gu2019}. In 2013, Gong \textit{et al.} considered the periodic auto-correlation of a single Golay sequence and proposed a systematic construction such that each of the sequence have a zero auto-correlation zone \cite{Gong2013}. This property plays a very important role in the synchronization and detection of signals. Recently in 2018, Chen \textit{et al.} \cite{Chen2018} further studied the zero periodic cross-correlation of the GCPs and proposed a new class of sequence sets, namely Golay-ZCZ sets. However, the aperiodic cross-correlation of the GCPs was never been considered till date.

Beside the theoretical approaches for the systematic constructions of the sequence pairs, numerous numerical approaches are also proposed till date, beginning with the work of Groot \textit{et al.} \cite{doi:10.1080/02331939208843771} in 1992. Groot \textit{et al.} related the problem of optimizing the merit factor of a sequence with thermodynamics and introduced some evolutionary strategies using optimization tools to get binary sequences having low auto-correlation. Since the merit factor of a sequence is highly multimodal (i.e., it
may have multiple local maxima) stochastic optimization algorithms had been used for its maximization \cite{doi:10.1080/02331939208843771}. However, for large values of $N$, the computational complexity of these algorithms become very high. To overcome this, Stoica \textit{et al.} made a remarkable progress and introduced several cyclic algorithms (CAs), namely CA-pruned (CAP) \cite{stoica2009new}, CA-new (CAN) \cite{stoica2009new}, and Weighted CAN (We-CAN) \cite{stoica2009new} to design sequences with low aperiodic autocorrelation. To generate the sequences with low periodic correlations, the authors also proposed periodic CAN (PeCAN) \cite{Stoica2009}. To design waveforms with arbitrarily spectral shapes, PeCAN has been modified as the SHAPE algorithm \cite{Rowe2014}. Inspired by the ideas of \cite{stoica2009new}, Soltanalian \textit{et al.} \cite{soltanalian2013fast} proposed a CAN algorithm for complementary sets and termed it as CANARY. In 2012, Soltanalian \textit{et al.} proposed a computational framework based on an iterative twisted approximation (ITROX) \cite{Soltanalian2012} and
a set of associated algorithms to generate sequences with good periodic/aperiodic correlation properties. In another work, to minimize the correlation magnitudes in desired intervals, in 2013, Li \textit{et al.} \cite{li2014waveform} proposed an approach based on iterative spectral approximation algorithm (ISAA) and derivative-based non-linear optimization algorithms. Gradient based algorithms were proposed in \cite{zhang2016cognitive,arlery2016efficient,arlery2016eo,tan2016phase} to design sequences with good correlation properties, in the frequency domain.

Meanwhile, in 2015, Song \textit{et al.} \cite{song2015optimization} proposed an algorithm to directly minimize the periodic/aperiodic correlation magnitudes at out-of-phase time-shifts, using the general majorization-minimization (MM) method. On the other hand Liang \textit{et al.} \cite{liang2016unimodular} proposed unimodular sequence design using alternating direction method of multipliers (ADMM). ADMM is a superior optimization technique as it  decomposes a
constrained convex optimization problem into multiple smaller
sub-problems whose solutions are coordinated to find the global
optimum. This form of decomposition-coordination procedure
allows parallel and/or distributed processing, and thus is well
suited to the handling big data. Second, in spite of employing iterations in the parameter updating process, it provides superior
convergence properties.

It should be noted that, designing of unimodular sequences using various optimization techniques is a decade old problem. However, the problem of designing complementary sequences considering the aperiodic cross-correlation among the sequences have not been considered before.


\subsection{Doppler resilient waveforms in polarimetric radar systems}
Fully polarimetric radar systems are equipped with vertically/
horizontally $(V/H)$ dual-dipole elements at every antenna
to make sure the simultaneous occurrence of transmitting and
receiving on two orthogonal polarizations \cite{calderbank2006instantaneous,pezeshki2008doppler,cui2017broadband}. The
essential ability of polarimetric radar systems is to capture the
scattering matrix which contains polarization properties of the
target. The scattering matrix can be given as follows,
\begin{equation}\label{eq:sm}
	\mathbf{H}=\begin{bmatrix}
	h_{VV} & h_{VH}\\
	h_{HV} & h_{HH}
	\end{bmatrix},
\end{equation}
where $h_{VH}$ denotes the target scattering coefficient that indicates the polarization change from $H$ (horizontally polarized incident field) into $V$ (vertical polarization channel).

The elements of the scattering matrix are estimated by analysing the auto-ambiguity functions (AAF) and cross-ambiguity functions (CAF) which are the matched filter outputs of the received signal with the transmitted
waveforms. Owing to its ideal ambiguity plot at desired delays waveforms with
impulse-like autocorrelation functions play an important role in radar applications. Phase coding is a commonly used to generate waveforms with impulse-like auto-correlations. Due to its ideal auto-correlation sum properties Howard \textit{et al.} \cite{howard2007simple} and Calderbank \textit{et al.} \cite{calderbank2006instantaneous} combined Golay complementary waveforms with Alamouti signal processing to enable pulse compression for multichannel and fully polarimetric radar systems.
One of the main drawbacks of waveforms phase coded with complementary sequences is that its effective ambiguity function is highly sensitive to Doppler shifts. Since then, several works have been done \cite{4339447,4784163,1405387} to design waveforms using sequences with good correlation properties which can exhibit some tolerance to Doppler shift. Working in this direction Pezeshki \textit{et al.} \cite{pezeshki2008doppler} made a remarkable progress in 2008, by designing Doppler resilient waveforms using Golay complementary sets. In \cite{pezeshki2008doppler} the transmission is determined by Alamouti coding and Prouhet-Thue-Morse (PTM) sequences. Extending further Tang \textit{et al.} \cite{tang2014construction} proposed Doppler resilient complete complementary code in multiple-input multiple-output (MIMO) radar by using generalised PTM sequences. Since, complementary sequences are not available for all lengths, in search of other sequences to design Doppler resilient waveforms, Wang \textit{et al.} \cite{Wang2017} proposed Z-complementary waveforms using equal sums of (like) powers (ESP) Sequences.

However, one of the major drawbacks of \cite{howard2007simple,calderbank2006instantaneous,pezeshki2008doppler} is that, the authors did not consider to design the dual-orthogonal waveforms for polarimetric radar systems. Although fully polarimetric radar systems simultaneously transmit and receive waveforms on two orthogonal polarizations, however, only this property does not help to extract the co- and cross-polarized scatter matrix elements \cite{224131,TITINSCHNAIDER2003633}. Therefore in polarimetric radar with simultaneous measurement of scattering matrix elements
the waveform need to have an extra orthogonality in addition to polarization orthogonality. Waveforms having two such orthogonality are called dual orthogonal waveforms.

\subsection{Contributions}
 One objective of this paper is to propose a novel class of sequence pairs, called ``quasi-orthogonal Z-complementary pairs (QOZCPs)", each depicting Z-complementary property for their aperiodic auto-correlation sums and also having a low correlation zone when their aperiodic cross-correlation is considered. Construction of QOZCPs based on Successively Distributed Algorithms under Majorization Minimization (SDAMM) are presented.
 Another objective of this paper is to apply the proposed QOZCPs in fully polarimetric radar systems and analyse the corresponding ambiguity functions. To be more precise, the contributions of this paper can be listed as follows:
\begin{enumerate}
	\item New pairs of sequences called QOZCPs are proposed.
	\item An efficient successively distributed algorithm under the MM framework is proposed. SDAMM transform a difficult optimization problem of two variables into two parallel sub-problems of a single variable. Each sub-problem has a closed-form so that the complexity is reduced significantly. Using SDAMM, we construct QOZCPs of any lengths.
	\item Extending the works of Pezeshki \textit{et al.} we propose a Doppler resilient dual orthogonal waveform based on QOZCPs which can be used to efficiently estimate the co- and cross- polarised scatter matrix elements.
	\item We compare the ambiguity plots of the proposed QOZCPs with the ambiguity plots of existing DR-GCPs and show that QOZCPs performs better while comparing the CAF plot.
\end{enumerate}

\subsection{Organization}
The rest of the paper is organised as follows. In section II, along with the preliminaries the definition of the QOZCP is proposed. The objective function for the construction of the proposed QOZCPs is derived. In Section III, SADMM algorithm is proposed to solve the optimization problem. In Section IV, we have explained the application of the proposed QOZCPs in radar waveform design. In section V, the numerical experiments are given, where we have compared the ambiguity plots of the proposed QOZCPs with the ambiguity plots of DR-GCPs. Finally, we have given some concluding remarks in section VI.

\section{QOZCP and Problem Formation}

In this section, we will propose the definition and the corresponding properties of QOZCPs. Throughout this paper, the entries of QOZCPs are complex $q$-th roots of unity (unimodular). Then we formally define QOZCPs as follows.

\subsection{Notations}
\begin{itemize}
	\item $\mathbf x^T$ and $\mathbf x^H$ denote the transpose and the Hermitian transpose of vector $\mathbf x$, respectively.
	\item $|x_l|$ and $x_l^*$ denote the modulus of $x_l$ and conjugate of $x_l$ respectively, where $x_l$ is the entries of $\mathbf x$.
	\item  $ \circ $ denotes the Hadamard product.
	\item $C_{xy} (k) $ denotes the aperiodic cross-correlation function of $\mathbf x=[x_0,x_1,\cdots, x_{L-1}]$ and $\mathbf y=[y_0,y_1,\cdots, y_{L-1}]$, i.e.,
	\begin{equation}
	C_{xy} (k) =\left\{
	\begin{aligned}
	\sum_{l=0}^{L-k-1}x_l y_{l+k}^*,\quad k\geq 0\\
	\sum_{l=0}^{L+k-1}x_{l-k}y_l^*,\quad k<0.\\
	\end{aligned}
	\right.
	\end{equation}
	\item $C_x (k) $ denotes the aperiodic auto-correlation function of $\mathbf x=[x_0,x_1,\cdots, x_{L-1}]$, i.e., $C_x(k) = C_{xx}(k)$.
	\item $X(z)$ denotes z-transform of $\mathbf x$, i.e., $X(z) =  x_0+x_1z^{-1}+\cdots+x_{L-1}z^{-(L-1)}$.
\end{itemize}

\begin{definition}
A pair of length-$L$ sequences $(\mathbf x,\mathbf y)$ is called a $(L,Z)$-QOZCP, if
\begin{eqnarray*}
\begin{split}
	&\textit{C}1: |C_x (k)+C_y(k)| \leq \epsilon, \mathrm{for\,\,  any}  \,\,   0<|k|<Z,\\
	&\textit{C}2: |C_{xy} (k)|\leq \epsilon, \mathrm{for\,\, any} \,\, |k|<Z,
\end{split}
\end{eqnarray*}

where $\epsilon$ is a very small positive real number which is very close to zero.
\end{definition}

Remark: since $C_{yx}(-k) = C_{xy}^*(k)$ and $ |C_{xy} (k)|\leq \epsilon$, then $|C_{yx}(k)|\leq \epsilon$ for any $|k|<Z$.

\begin{figure}[htbp]
\centering
\subfigure[$|C_x(\tau)+C_y(\tau)|$]{
\begin{minipage}{7cm}
\centering
\includegraphics[width=7cm]{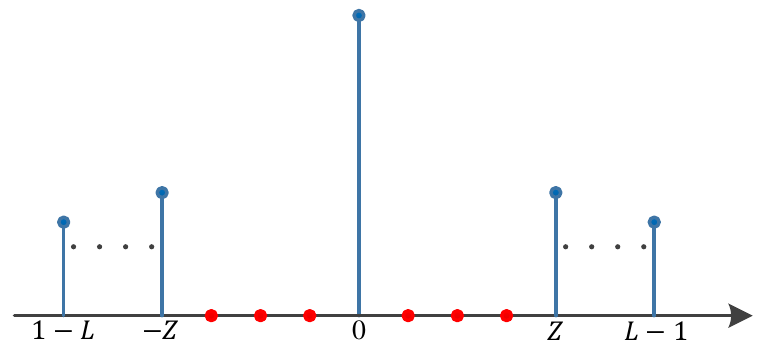}\\
\end{minipage}
}
\subfigure[$|C_{xy}(\tau)|$]{
\begin{minipage}{7cm}
\centering
\includegraphics[width=7cm,height = 3.4cm]{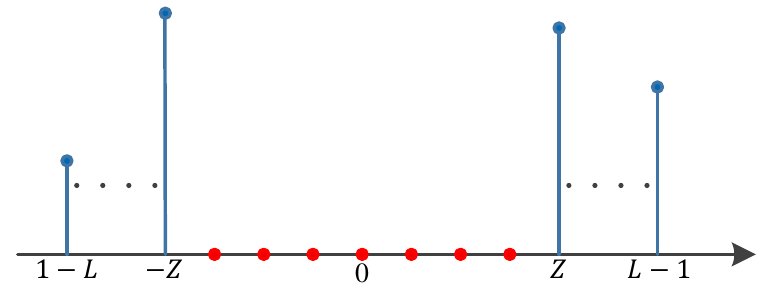}
\end{minipage}
}
\caption{ Illustrative plots for the correlation properties of (L,Z)-QOZCP}
\label{fig: C}
\end{figure}

According to Definition 1, it can be observed that each QOZCP has low zero correlation zones when the ACFs' sum and the CCF are considered. It is noted that the nonzero values in the low zero autocorrelation zone are very close to zero. We illustrate the correlation properties of $(L,Z)$-QOZCP with $\epsilon=0$ in Fig. \ref{fig: C}. Following this definition of sequence, the task of constructing QOZCPs is transformed into equivalent optimization problems. The optimization problem will be formulated after introducing the objective function and constraints.

\subsection{Objective Function}
We have already introduced the desired sequence pair. In order to find it,  unified metrics named the Weighted Complementary Integrated Sidelobe Level (WCISL) and the Weighted Cross-Correlation Integrated Level (WCCIL) are proposed in the following definitions.

\emph{Definition 2:} The Weighted Complementary Integrated Sidelobe Level (WCISL)  of a sequence pair $(\mathbf x, \mathbf y)$  is defined as
\begin{equation}
\mathrm{WCISL} = \sum_{k=1}^{L-1}w_k|C_x (k)+C_y (k)|^2.
\end{equation}
\emph{Definition 3:} The weighted cross-correlation integrated level(WCCIL) of $\mathbf x$, $\mathbf y$ is given by
\begin{equation}
\mathrm{WCCIL} =\sum_{k=0}^{L-1 }\tilde{w}_k | {C_{xy}}(k)|^2,
\end{equation}
where $C_{x}(k)$ and $C_{xy}(k)$ are the auto-correlation function of $\mathbf x$ and the cross-correlation function of $\mathbf x,\mathbf y$, respectively. Besides, $w_{-k}=w_k$, $w_0 = 0$, $\tilde{w}_{-k}=w_k$, $\tilde{w}_0 \ne 0$.

The expressions of WCISL and WCCIL can also be transformed into the following expressions,
\begin{equation}
\begin{split}
&\mathrm{WCISL} =\sum_{k=1}^{L-1} w_k( |\mathbf x^H \mathbf{U}_k \mathbf x+\mathbf y^H \mathbf{U}_k \mathbf y|^2),\\
&\mathrm{WCCIL} =\sum_{k=0}^{L-1} w_k |\mathbf x^H \mathbf{U}_k \mathbf y|^2,
\end{split}
\end{equation}
where $\mathbf{U}_{k}$ \cite{song2016sequence} is an $N\times N$ Toeplitz matrix with 1 in $k$-th diagonal and 0 in the other positions.

To satisfy the conditions, i.e., C1 and C2 at the same time, WCISL and WCCIL should be considered together.
Then the objective function is written as
\begin{equation}
\begin{split}
\alpha \sum_{k=1}^{L-1} w_k( |\mathbf x^H \mathbf{U}_k \mathbf x&+\mathbf y^H \mathbf{U}_k \mathbf y|^2) +(1-\alpha)\sum_{k=0}^{L-1} w_k |\mathbf x^H \mathbf{U}_k \mathbf y|^2,\label{obj: 1}
\end{split}
\end{equation}
which will be minimized within a constraint set.

\subsection{Constraints of Interest}
Usually, the sequences to be designed have limited energy \cite{zhao2016unified}. In addition, the large PAPR results in a difficult dilemma between power efficiency and signal distortion \cite{wang2018optimized}. Therefore, constraints of energy and PAPR should be considered as follows:

1) Energy Constraint: The energy of $\mathbf x$ and $\mathbf y$ should be constrained to a given power $p_e$, i.e.,
\begin{equation}
||\mathbf x||^2=p_e, ||\mathbf y||^2=p_e, \label{con: pe}
\end{equation}
where $p_e$ is not larger than $L$(L is the length of $\mathbf x$ or $\mathbf y$).

2) PAPR Constraint: The PAPR is the ratio of the largest signal magnitude to its average power \cite{tropp2005designing,zhao2016unified}:
\begin{equation}
\mathrm{PAPR}(\mathbf x) = \frac{\max_{l}|x_l |^2}{||x||^2/L} \mathrm{and}\,\, \mathrm{PAPR}(\mathbf y) = \frac{\max_{l}|y_l |^2}{||\mathbf y||^2/L},
\end{equation}
where $1\leq \mathrm{PAPR}(\mathbf x)\leq L$ and $1\leq \mathrm{PAPR}(\mathbf y)\leq L$. We require a threshold $p_r(<L)$ which is determined by  power amplifiers of the system, and let $\mathrm{PAPR}(x)<p_r$ and $\mathrm{PAPR}(x)<p_r$, so that the sequences can be with high  power efficiency and small signal distortion. Since we have already set $||\mathbf x||^2=p_e$ and $||\mathbf y||^2=p_e$, the PAPR constraints are equivalent to: for $l = 1,2,\cdots, L$,
\begin{equation}
|x_l|\leq p_c,\, |y_l|\leq p_c,\label{con: pc}
\end{equation}
where $p_c = \sqrt{p_r p_e/L}$.

\subsection{Problem Formulation}
The problem formation is composed of the minimization of the objective shown in (\ref{obj: 1}) subject to the constraints (\ref{con: pe}) (\ref{con: pc}), and it reads
\begin{equation}
\mathcal{P}_0\left\{
\begin{aligned}
\begin{split}
\min \limits_{\mathbf x,\mathbf y}\quad &\alpha \sum_{k=1}^{L-1} w_k( |\mathbf x^H \mathbf{U}_k \mathbf x+\mathbf y^H \mathbf{U}_k \mathbf y|^2) +(1-\alpha)\sum_{k=0}^{L-1} w_k |\mathbf x^H \mathbf{U}_k \mathbf y|^2,\label{obj: 1}\\
s.t.\quad & ||\mathbf x||^2=p_e, ||\mathbf y||^2=p_e;\\
& |x_l|\leq p_c, |y_l|\leq p_c, \, \mathrm{for}\,\, l=0,1,2,\cdots, L-1.
\end{split}
\end{aligned}
\right.
\end{equation}

Besides, if the $\mathrm{PAPR}(x)$ and $\mathrm{PAPR}(y)$ are equal to 1,  the optimization problem $\mathcal{P}_0$ can be changed into  $\mathcal{P}_1$:
\begin{equation}
\mathcal{P}_1\left\{
\begin{aligned}
\begin{split}
\min \limits_{\mathbf x,\mathbf y}\quad &\alpha \sum_{k=1}^{L-1} w_k( |\mathbf x^H \mathbf{U}_k \mathbf x+\mathbf y^H \mathbf{U}_k \mathbf y|^2) +(1-\alpha)\sum_{k=0}^{L-1} w_k |\mathbf x^H \mathbf{U}_k \mathbf y|^2,\label{obj: 1}\\
s.t.\quad & | x_l |=1, | y_l |=1,   \mathrm{for}\,\, l=0,1,2,\cdots, L-1.\\
\end{split}
\end{aligned}
\right.
\end{equation}

The optimization problems $\mathcal{P}_0$ and $\mathcal{P}_1$ are difficult to solve, since:\\
1) The objective function is non-convex and quartic for $\mathcal{P}_0$ and $\mathcal{P}_1$.\\
2) The variables $\mathbf x$ and $\mathbf y$ are difficult to be separated.\\
3) The constraint set is not a convex set.

In the following, we will pay more attention to dealing with $\mathcal{P}_0$, and later solve $\mathcal{P}_0$ without extra effort.

\section{QOZCP Design under Majorization Minimization Framework}
In this section, we will propose a Successively Distributed Algorithms under the Majorization Minimization (SDAMM) framework to solve $\mathcal{P}_0$ and $\mathcal{P}_1$.

\subsection{SDAMM for $\mathcal{P}_0$}

In order to analyze the proposed optimization problem conveniently, we combine the two optimization variables $\mathbf x$, $\mathbf y$  into one variable, i.e., $\mathbf{z} = [\mathbf{x}^T , \mathbf{y}^T ]^T$.
Then the optimization problem can be changed into
\begin{equation}
\begin{split}
\begin{aligned}
\min \limits_{\mathbf z,\mathbf x, \mathbf y}&\quad \alpha \sum_{k=-L+1}^{L-1} w_k |{\mathbf  z^H \mathbf  A_k \mathbf  z}|^2+(1-\alpha)\!\! \sum_{k=-L+1}^{L-1} \tilde{w}_k  | {\mathbf z^H \mathbf B_k \mathbf z}|^2\\
s.t.  &\,\,  \mathbf z = [\mathbf x^T, \mathbf y^T]^T\\
& ||\mathbf x||^2=p_e, ||\mathbf y||^2=p_e\\
& |x_l|\leq p_c, |y_l|\leq p_c, \, \mathrm{for}\,\, l=0,1,2,\cdots, L-1,
\end{aligned}\label{opt:p00}
\end{split}
\end{equation}
where
$\mathbf {A_k} = \begin{bmatrix} \mathbf U_k&  \mathbf O\\  \mathbf O& \mathbf U_k \end{bmatrix},\quad \mathbf{B_k} = \begin{bmatrix}  \mathbf O& \mathbf U_k\\ \mathbf O& \mathbf O \end{bmatrix}.$

\begin{proposition}
The optimization problem (\ref{opt:p00}) can be transformed as
\begin{equation}
\mathcal{P}_{0,1}\left\{
\begin{aligned}
\min \limits_{\mathbf Z,\mathbf z,\mathbf x, \mathbf y}&\quad \text{vec}(\mathbf Z)^H \mathbf J \text{vec}(\mathbf Z)\\
s.t.& \quad \mathbf Z= \mathbf {zz}^*\\
 &\quad  ||\mathbf x||^2=p_e, ||\mathbf y||^2=p_e;\\
& \quad |x_l|\leq p_c, |y_l|\leq p_c, \, \mathrm{for}\,\, l=0,1,2,\cdots, L-1.
\end{aligned}\label{opt:p01}
\right.
\end{equation}
where $\mathbf J= \alpha \mathbf J_A+(1-\alpha)\mathbf J_B$,
\begin{eqnarray} \mathbf{J}_{A} =\sum_{k=-L+1}^{L-1} w_{k} \operatorname{vec}\left(\mathbf{A}_{k}\right) \operatorname{vec}\left(\mathbf{A}_{k}\right)^{H},  \mathbf{J}_{B} =\sum_{k=-L+1}^{L-1} w_{k} \operatorname{vec}\left(\mathbf{B}_{k}\right) \operatorname{vec}\left(\mathbf{B}_{k}\right)^{H}. \end{eqnarray}

\end{proposition}
\begin{IEEEproof}
Please see Appendix A
\end{IEEEproof}

Now, we can use the framework of MM to dispose of the problem.

\begin{proposition}
Optimization problem $\mathcal{P}_{0,1}$ can be majorized by the following problem at $\mathbf z^{(t)}$
\begin{equation}
\mathcal{P}_{0,2}\left\{
\begin{aligned}
\min \limits_{\mathbf Z,\mathbf z,\mathbf x, \mathbf y}&\quad \text{Re}\left\{\text{vec}(\mathbf Z)^H (\mathbf J-\lambda_{\mathbf J}\mathbf I) \text{vec}(\mathbf Z^{(t)})\right\}\\
s.t.& \, \mathbf Z= \mathbf {zz}^*\\
 &\,\,  \mathbf z = [\mathbf x^T, \mathbf y^T]^T\\
& ||\mathbf x||^2=p_e, ||\mathbf y||^2=p_e\\
& |x_l|\leq p_c, |y_l|\leq p_c, \, \mathrm{for}\,\, l=0,1,2,\cdots, L-1,
\end{aligned}\label{opt:p02}
\right.
\end{equation}
where $\lambda_{\mathbf J}$ is the largest eigenvalue of $\mathbf J$.
\end{proposition}
\begin{IEEEproof}
Please see Appendix B
\end{IEEEproof}

\begin{theorem}
$\lambda_{\mathbf J} $, the largest eigenvalue of $\mathbf J$, is equal to
\begin{equation}
\begin{split}
&\lambda_{\mathbf J} = \max_k \left\{\max\{\lambda_{A}(k), \lambda_{B}(k) \right\}|k=-L+1,\cdots,L-1\},
\end{split}
\end{equation}
where
\begin{equation}
\begin{split}
&\lambda_{A}(k) = w_k \alpha (2N-2|k|); \lambda_{B}(k) = w_k (1-\alpha) (N-|k|).
\end{split}
\end{equation}
\end{theorem}
\begin{IEEEproof}
Please see Appendix C.
\end{IEEEproof}

\begin{proposition}
Optimization problem $\mathcal{P}_{0,2}$ can be transformed into the following problem

\begin{equation}
\mathcal{P}_{0,3}\left\{
\begin{aligned}
\min \limits_{\mathbf z,\mathbf x, \mathbf y}&\quad \text{Re}\left\{\mathbf z^H \left( \mathbf Q -\lambda_{max}(\mathbf J) \mathbf z^{(t)}(\mathbf z^{(t)})^H\right)\mathbf z\right\}\\
s.t. &\,\,  \mathbf z = [\mathbf x^T, \mathbf y^T]^T\\
& ||\mathbf x||^2=p_e, ||\mathbf y||^2=p_e\\
& |x_l|\leq p_c, |y_l|\leq p_c, \, \mathrm{for}\,\, l=0,1,2,\cdots, L-1,
\end{aligned}\label{opt:p03}
\right.
\end{equation}
 where
 \begin{eqnarray}
 \begin{split}
&\mathbf {Q}
= \begin{bmatrix} \alpha\!\!\displaystyle \sum_{k=-L+1}^{L-1} \!\! w_k r_{-k}^{(l)}\mathbf U_k&\quad  (1\!-\!\alpha)\!\displaystyle\sum_{k=-L+1}^{L-1}  w_k c_{-k}^{(l)}\mathbf U_k\\
 \mathbf O&\!\! \alpha\displaystyle \sum_{k=-L+1}^{L-1}  w_k r_{-k}^{(l)}\mathbf U_k \end{bmatrix},
\end{split}\label{eq:Q}
\end{eqnarray}
in which $r_{-k}=C_x (-k)+C_y (-k)$ and $c_{-k}=C_{xy}(-k)$,
\end{proposition}

\begin{IEEEproof}
Please see Appendix D
\end{IEEEproof}

There are many operations in $\mathbf Q$ shown in (\ref{eq:Q}). In order to decrease the complexity of computing $\mathbf Q$, FFT (IFFT) is used.

\begin{theorem}
 $r_{k}^{(l)}$ and $c_{k}$ in $\mathbf Q$ can be computed by FFT(IFFT) operations as follows, respectively,
\begin{eqnarray}
\begin{split}
&\mathbf r = [r_0^{(l)},r_1^{(l)},\cdots,r_{L-1}^{(l)},0,r_{1-L}^{(l)},\cdots,r_{-1}^{(l)},]\\
&= \mathbf F^H\!|\mathbf F[(\mathbf x^{(t)})^T\!,\mathbf 0_{1\times L}]^T|^2\!+\!\mathbf F^H\!|\mathbf F[(\mathbf y^{(t)})^T\!,\mathbf 0_{1\times L}]^T|^2,
\end{split}
\end{eqnarray}
and
\begin{eqnarray}
  \begin{split}
 &\mathbf c = [c_0^{(l)},c_1^{(l)},\cdots, c_{L-1}^{(l)},0,c_{1-L}^{(l)},\cdots,c_{-1}^{(l)}]^T\\
 &= \mathbf F^H \left((\mathbf F[(\mathbf x^{(t)})^T,\mathbf 0_{1\times L}]^T)^*\circ (\mathbf F[(\mathbf y^{(t)})^T,\mathbf 0_{1\times L}]^T)\right),
 \end{split}
\end{eqnarray}
where $\mathbf F$ is a $2L\times2L$ discrete Fourier matrix whose element is $F_{il} = e^{-j2\pi\omega_i l/L}$ and $|\cdot|^2$ denotes the element-wise absolute-squared operation.
\end{theorem}
\begin{IEEEproof}
Please see Appendix E.
\end{IEEEproof}

According to Lemma 4 in \cite{song2016sequence} and some simple operations, we have the following equations,
\begin{equation}
\begin{split}
&\mathbf{Q} = \left[\begin{matrix}
\frac{\alpha}{2L}\mathbf F_{:,1:L}^H diag(\mathbf \mu_r) \mathbf F_{:,1:L}& \frac{1-\alpha}{2L}\mathbf F_{:,1:L}^H Diag(\mathbf \mu_c)\mathbf F_{:,1:L}\\
\mathbf O & \frac{\alpha}{2L}\mathbf F_{:,1:L}^H diag(\mathbf \mu_r) \mathbf F_{:,1:L}
\end{matrix}\right],\label{Q}
\end{split}
\end{equation}

where $\mathbf {\mu_r = F t_r}$, $\mathbf \mu_c = \mathbf F \mathbf t_c$, $\mathbf f_x = \mathbf F[(\mathbf x^{(t)})^T,\mathbf 0_{1\times L}]^T $, $\mathbf f_y = \mathbf F[(\mathbf y^{(t)})^T,\mathbf 0_{1\times L}]^T $,
 \begin{equation}
 \begin{split}
 &\mathbf {t_r} = [0,w_1 r_1^{(l)},\cdots, w_{L-1}r_{L-1}^{(l)},0,w_{L-1}r_{1-L}^{(l)},\cdots,w_1 r_{-1}^{(l)}],\\
 &\mathbf {t_c} = [w_0 c_0^{(l)}\!\!,w_1 c_1^{(l)}\!\!,\cdots\!\!, w_{L-1}c_{L-1}^{(l)}, 0, w_{L-1}c_{1-L}^{(l)},\!\cdots\!\!,w_1 c_{-1}^{(l)}].
 \end{split}
\end{equation}


Since $\mathbf{Q} -\lambda_{max}(\mathbf J) \mathbf z^{(t)}$ in (\ref{opt:p03}) is not Hermitian, the optimization problem  (\ref{opt:p03}) is not  a traditional Unimodular Quadratic Programming (UQP) defined in \cite{soltanalian2014designing}. Then we have to address the problem and transform (\ref{opt:p03}) into a UQP in next theorem.
\begin{proposition}
 The optimization problem $\mathcal{P}_{0,3}$ can be equivalently transformed into the following optimization problem:
\begin{equation}
\mathcal{P}_{0,4}\left\{
\begin{aligned}
\min \limits_{\mathbf z,\mathbf x,\mathbf y}&\quad \mathbf z^H \left( \mathbf Q +\mathbf Q^H-2\lambda_{\mathbf J} \mathbf z^{(t)}(\mathbf z^{(t)})^H\right)\mathbf z\\
s.t. &\,\,  \mathbf z = [\mathbf x^T, \mathbf y^T]^T\\
& ||\mathbf x||^2=p_e, ||\mathbf y||^2=p_e\\
& |x_l|\leq p_c, |y_l|\leq p_c, \, \mathrm{for}\,\, l=0,1,2,\cdots, L-1.
\end{aligned}\label{opt:p04}
\right.
\end{equation}
\end{proposition}
\begin{IEEEproof}
 We can transform (\ref{opt:p03}) into  a UQP by adding a conjugate term $\mathbf{Q}^H-\lambda_{\mathbf{J}}\mathbf{z}^{(t)}(\mathbf{z}^{(t)})^H$ to the objective of (\ref{opt:p03}), then the objective function of UQP is written as
\begin{equation}
\begin{aligned}
 \text{Re}\left\{\mathbf z^H \left( \mathbf Q +\mathbf Q^H-2\lambda_{\mathbf J} \mathbf z^{(t)}(\mathbf z^{(t)})^H\right)\mathbf z \right\}\\
\end{aligned}
\end{equation}
whose result of optimal variable is not changed compared with (\ref{opt:p03}).
Also, the operation $\text{Re}(\cdot)$ can be removed, because $\left( \mathbf Q +\mathbf Q^H-2\lambda_{\mathbf J} \mathbf z^{(t)}(\mathbf z^{(t)})^H\right)$ has been Hermitian and
\begin{eqnarray}
	\mathbf z^H\left( \mathbf Q +\mathbf Q^H-2\lambda_{\mathbf J} \mathbf z^{(t)}(\mathbf z^{(t)})^H\right)\mathbf z
\end{eqnarray}
is a real number. Then the optimization problem can be changed into $\mathcal{P}_{0,4}$.
\end{IEEEproof}

\begin{proposition}
 The optimization problem $\mathcal{P}_{0,4}$ can be majorized by the majorization problem at $\mathbf z^{(t)}$:
\begin{equation}
\mathcal{P}_{0,5}\left\{
\begin{aligned}
\min \limits_{\mathbf z}&\quad \text{Re}\left\{\mathbf z^H \left( \mathbf Q +\mathbf Q^H-2\lambda_{\mathbf J} \mathbf z^{(t)}(\mathbf z^{(t)})^H-\lambda_u\right)\mathbf z^{(t)} \right\}\\
s.t. &\,\,  \mathbf z = [\mathbf x^T, \mathbf y^T]^T\\
& ||\mathbf x||^2=p_e, ||\mathbf y||^2=p_e\\
& |x_l|\leq p_c, |y_l|\leq p_c, \, \mathrm{for}\,\, l=0,1,2,\cdots, L-1,
\end{aligned}
\right.
\end{equation}
where
$\lambda_u = 4L (\max \limits_{1\le i,j\le L}|\mathbf Q_{i,j}|)$, $\lambda_{\mathbf J}$ is the largest eigenvalue of $\mathbf J= \alpha \mathbf J_A+(1-\alpha)\mathbf J_B$.
\end{proposition}
\begin{IEEEproof}
Please see Appendix F.
\end{IEEEproof}

The objective of $\mathcal{P}_{0}$ can be majorized by the objective of $\mathcal{P}_{0,5}$ at $\mathbf z^{(t)}$, i.e., $-\text{Re}\{\mathbf z^H \mathbf P(\mathbf z^{(t)})\}$,
where
\begin{eqnarray}
\begin{split}
 \mathbf P({\mathbf z^{(t)}}) &= -(\mathbf Q+\mathbf Q^H-2 \lambda_{\mathbf J} \mathbf z^{(t)}(\mathbf z^{(t)})^H-\lambda_u \mathbf I)\mathbf z^{(t)}\\
 &=  (\lambda_{\mathbf J}\cdot 4L+\lambda_u)\mathbf z^{(t)}-(\mathbf Q+\mathbf Q^H) \mathbf z^{(t)},
 \end{split}
 \end{eqnarray}
 and
 \begin{eqnarray}
 \begin{split}
&(\mathbf Q+\mathbf Q^H) \mathbf z^{(t)} = \left[\begin{matrix} \frac{\alpha}{2L}\mathbf F_{:,1:L}^H ((\mathbf \mu_r+\bar{\mathbf \mu}_r) \circ \mathbf f_x)+\frac{1-\alpha}{2L}\mathbf F_{:,1:L}^H  (\mathbf \mu_c\circ \mathbf f_y)\\
 \frac{\alpha}{2L}\mathbf F_{:,1:L}^H ((\mathbf \mu_r+\bar{\mathbf \mu}_r) \circ \mathbf f_y)+\frac{1-\alpha}{2L}\mathbf F_{:,1:L}^H  ( \bar{\mathbf \mu}_c\circ \mathbf f_x)\end{matrix}\right].
 \end{split}\label{eq:qqz}
\end{eqnarray}

Besides,
\begin{eqnarray*}
\text{Re}\{\mathbf z^H \mathbf P(\mathbf z^{(t)}) \} = \text{Re}\{\mathbf x^H \mathbf P_x(\mathbf z^{(t)})\} +\text{Re}\{\mathbf y^H \mathbf P_y(\mathbf z^{(t)})\} ,
\end{eqnarray*}
where $\mathbf P_x(\mathbf z^{(t)}) = \mathbf P(\mathbf z^{(t)})(1:L)$ and $\mathbf P_y(\mathbf z^{(t)})=\mathbf p(\mathbf z^{(t)})(L+1:2L)$

In other words, $\mathcal{P}_0$ can be majorized by the following optimization at $\mathbf z^{(t)}$, i.e.,
\begin{eqnarray}
\mathcal{P}_{0,6}\left\{
\begin{aligned}
\begin{split}
\min \limits_{\mathbf x,\mathbf y}\quad &-\text{Re}\{\mathbf x^H \mathbf P_x(\mathbf z^{(t)})\}-\text{Re}\{\mathbf y^H \mathbf P_y(\mathbf z^{(t)})\}\\
s.t.\quad & ||\mathbf x||^2=p_e, ||\mathbf y||^2=p_e;\\
&|x_l|\leq p_c,|y_l|\leq p_c, \, \mathrm{for}\,\, l=0,1,2,\cdots, L-1.
\end{split}
\end{aligned}
\right.
\end{eqnarray}

The variables $\mathbf x$ and $\mathbf y$ are obviously separate  in $\mathcal{P}_{0,6}$. Therefore,  $\mathcal{P}_{0,6}$ can be divided into two optimization problems $\mathcal{P}_{0,x}$ and $\mathcal{P}_{0,y}$
\begin{eqnarray}
	\mathcal{P}_{0, x}\left\{\begin{array}{cl}\max\limits_{\mathbf x} & \operatorname{Re}\left\{\mathbf{x}^{H} \mathbf{P}_{x}\left(\mathbf{z}^{(t)}\right)\right\} \\ \text { s.t. } & \|\mathbf{x}\|^{2}=p_{e}, \\ & \left|x_{l}\right| \leq p_{c},\end{array},
	\quad \mathcal{P}_{0, y}\left\{\begin{array}{cc}\max \limits_{\mathbf{y}} & \operatorname{Re}\left\{\mathbf{y}^{H} \mathbf{P}_{y}\left(\mathbf{z}^{(t)}\right)\right\} \\ \text {s.t.} & \|\mathbf{y}\|^{2}=p_{e}, \\ & \left|y_{k}\right|^{2} \leq p_{c}.\end{array}\right.\right.
\end{eqnarray}

This kind of problem like $\mathcal{P}_{0,x}$ or $\mathcal{P}_{0,y}$ has been solved in \cite{yang2018cognitive} with a closed form denoted as $\mathbf x = \mathrm{Proj}_0 (\mathbf P_x(\mathbf z^{(t)})$ which is shown in Appendix G.

Then the proposed algorithm based on MM framework for $\mathcal{P}_0$ is summarized in Algorithm 1.

\subsection{SDAMM for $\mathcal{P}_1$}
Since $\mathcal{P}_1$ is a special case of $\mathcal{P}_0$ and  the objective function of $\mathcal{P}_1$ is the same with that of $\mathcal{P}_0$, we can directly use the objective functions of $\mathcal{P}_{0,x}$ and $\mathcal{P}_{0,y}$ as the majorized functions of $\mathcal{P}_1$ at $\mathbf x^{(t)}$ and $\mathbf y^{(t)}$, respectively. Therefore, the two optimization problems are

\begin{eqnarray}
\mathcal{P}_{1,x}\left\{
\begin{aligned}
\max \limits_{\mathbf x}\quad &\text{Re}\{\mathbf x^H \mathbf P_x(\mathbf z^{(t)})\}\\
s.t.\quad &|x_l| = 1,
\end{aligned}
\right.
\quad
\mathcal{P}_{1,y}\left\{
\begin{aligned}
\begin{split}
\max \limits_{\mathbf y}\quad &\text{Re}\{\mathbf y^H\mathbf P_y(\mathbf z^{(t)})\}\\
s.t.\quad & |y_l|= 1.
\end{split}
\end{aligned}
\right.
\end{eqnarray}

According to \cite{soltanalian2014designing}, the minimizer $\mathbf x$ of $\mathcal{P}_{1,x}$ can be the equivalent minimizer of  the following optimization problem (\ref{opt:p8})
  \begin{equation}
\begin{aligned}
\min \limits_{\mathbf x}&\quad ||\mathbf x- \mathbf P_x({\mathbf z^{(t)}})||\\
s.t. &\quad |x_i| =1,\quad  i=1,2,\ldots,2L,  \\
\end{aligned}\label{opt:p8}
\end{equation}
similarly for $\mathbf y$.

 It is obvious that the problem (\ref{opt:p8}) has a closed form
 \begin{eqnarray}
\mathbf x = e^{arg(\mathbf P_x(\mathbf z^{(t)}))},
 \end{eqnarray}
where $arg(\cdot)$ represents the argument. We denote $\mathbf x = e^{arg(\mathbf P_x(\mathbf z^{(t)}))}$ as $\mathbf x = \mathrm{Proj}_1 (\mathbf P_x(\mathbf z^{(t)}))$. Also, $\mathbf y=\mathrm{Proj}_1(\mathbf P_y(\mathbf z^{(t)}))=e^{arg(\mathbf P_y(\mathbf z^{(t)}))}$.

Then the proposed algorithm based on MM framework is summarized in Algorithm 1.

\begin{algorithm}[htb]
\caption{SDAMM: Successively Distributed Algorithms under Majorization Minimization for $\mathcal{P}_i$, $i=0,1$:}
\label{alg:Framwork}
\begin{algorithmic}[1]
\REQUIRE ~~\\
sequence length $L$;\\
weights $\{w_k \ge 0 \}_{k=1}^{L-1}$, $\{\tilde{w}_k\ge 0\}_{k=0}^{L-1}$;\\
scalar $\alpha=\frac{1}{2}$.
\STATE Set $l=0$, initialize $\mathbf z^{(0)}=[(\mathbf x^{(0)})^T, (\mathbf y^{(0)})^T]^T$.
\label{ code:fram:extract }
\STATE $\lambda_A (k) = w_k\alpha (2L-2|k|)$, $ \lambda_B (k) =  w_k (1-\alpha)(L-|k|)$,\\
 $\lambda_{\mathbf J} = \max_k\left\{ \max(\lambda_A (k), \lambda_B (k) )| k=1,\cdots, L  \right\}$.
\label{code:fram:trainbase}
\STATE {\bf repeat}
\STATE $\mathbf x_1 = \mathrm{Proj}_i(\mathbf P_x (\mathbf z^{(t)}))$, $\mathbf y_1 = \mathrm{Proj}_i(\mathbf P_y (\mathbf z^{(t)}))$, \\
$\mathbf z_1 = [\mathbf x_1^T, \mathbf y_1^T]^T$
\STATE $\mathbf x_2 = \mathrm{Proj}_i(\mathbf P_x (\mathbf z^{(t)}))$, $\mathbf y_2 = \mathrm{Proj}_i(\mathbf P_y (\mathbf z^{(t)}))$,\\
$\mathbf z_2 = [\mathbf x_2^T, \mathbf y_2^T]^T$
\STATE $\mathbf v_1 = \mathbf z_1-\mathbf z^{(t)}$, $\mathbf v_2 = \mathbf z_2-\mathbf z_1- \mathbf v_1$
\STATE Compute the step length $\alpha_{sl}=-\frac{||\mathbf v_1||}{||\mathbf v_2||}$;
\STATE $\mathbf z^{(l+1)} =\mathrm{Proj}_i{(\mathbf P(\mathbf z^{(t)}-2\alpha_{sl}\mathbf v_1 +\alpha_{sl}^2 \mathbf v_2 ))}$
\STATE {\bf while} $\mathrm{obj}(\mathbf z)>\mathrm{obj}(\mathbf z^{(t)})$\\
 {\bf do}\\
\quad$\alpha_{sl} = (\alpha_{sl}-1)/2$,\\
\quad $\mathbf z^{(l+1)} =\mathrm{Proj}_i{(\mathbf P(\mathbf z^{(t)}-2\alpha_{sl}\mathbf v_1 +\alpha_{sl}^2 \mathbf v_2 ))}$\\
 end while
\STATE $l\leftarrow l+1$
\STATE {\bf{until}} convergence.
\end{algorithmic}
\end{algorithm}

 \section{Applications of QOZCP to Radar Waveform Design}
\subsection{QOZCPs for SISO Radar Waveform}
Suppose from a transmit antenna is transmitted a sequence vector $\mathbf s^T=[\mathbf s_0, \mathbf s_1,\cdots,\mathbf s_{N-2},\mathbf s_{N-1}]$, where $\{\mathbf s_n\}_{n=0}^{N-1}$ are length-$L$ sequences over $N$ pulse repetition intervals (PRIs).  It is noted that the sequences $\{\mathbf s_n\}_{n=0}^{N-1}$ are formed from QOZCP $(\mathbf x, \mathbf y)$ and their variants $\pm\mathbf x, \pm\mathbf y, \pm \tilde{\mathbf x}, \pm\tilde{\mathbf y}$, where $\tilde{\mathbf x}=(\tilde{x}[0], \tilde{x}[1], \cdots, \tilde{x}[L-1])$,   and $\tilde{\cdot}$ denotes reversed complex conjugate, i.e., $\tilde{x}[l]=x^*[-l]$  for $l=0,1,\cdots,L-1$.    Let $S_n (z) = \mathcal{Z}\{\mathbf s_n\}$  be the $z$-transform of $\mathbf s_n$ so that
\begin{eqnarray}
S_n(z) = s_n [0]+s_n [1]z^{-1}+\cdots+s_n [L-1]z^{-(L-1)}
\end{eqnarray}
for $n=0,1,\cdots,N-1$. Then $\mathbf S(z)$, the transmit sequence vector in $z$-domain, is written as
 \begin{eqnarray}
 \mathbf S^T(z)=[S_0(z),S_1(z),\cdots,S_{N-2}(z),S_{N-1}(z)].
 \end{eqnarray}

As the assumption in \cite{pezeshki2008doppler}, the scatterer with a constant velocity has equal intra-Doppler in every PRI, whereas it has a relative Doppler shift $\theta_0$ between adjacent PRIs.
Then,  in $n$-th PRI, $R_n(z)$, the returned sequence in $z$-domain, associate with a scatter at delay coordinate $d_0$, is denoted as
\begin{eqnarray}
R_n(z) = h_0 z^{-d_0}S_n(z) e^{jn\theta_0}+W_n(z)
\end{eqnarray}
where $h_0$ is a scattering coefficient and $w_n (z)$ is a noise in $n$-th PRI.
The returned sequence vector in $z$-domain is written as
\begin{eqnarray}
\mathbf R^T(z) = h_0 z^{-d_0}\mathbf S^T (z) \mathbf{D}(\theta_0)+\mathbf W^T(z),
\end{eqnarray}
where $\mathbf R^T(z) = [R_0 (z), R_1(z),\cdots, R_{N-2}(z), R_{N-1}(z)]$ , $\mathbf W^T(z)=[W_0 (z), W_1(z),\cdots, W_{N-2}, W_{N-1}]$, and $\mathbf{D}(\theta)$ is a diagonal Doppler modulation matrix which can be denoted as
\begin{eqnarray}
\mathbf{D}(\theta) = \mathrm{diag}(1,e^{j\theta},\cdots,e^{j(N-1)\theta}).
\end{eqnarray}

The received sequence vector $\mathbf R^T(z)$ is processed by the receiver vector $\tilde{\mathbf S}(z)$, i.e.,
\begin{eqnarray}
\tilde{\mathbf S}(z) = [\tilde{S}_0(z),\tilde{S}_1(z),\cdots,\tilde{S}_{N-2}(z),\tilde{S}_{N-1}(z)]^T
\end{eqnarray}
where $\tilde{S}_n(z)=S^*(1/z^*)$ is the $z$-transform of $\tilde{\mathbf s}_n$.

Then the output is written as
\begin{eqnarray}
\mathbf R^T(z) \tilde{\mathbf S}(z) = h_0 z^{-d_0} G(z,\theta_0)+\mathbf W^T(z) \tilde{\mathbf S}(z),
\end{eqnarray}
where $G(z,\theta)$ is given by
\begin{eqnarray}
G(z,\theta) = \mathbf S(z) \mathbf{D}(\theta) \mathbf S(z) = \sum_{n=0}^{N-1}e^{jn\theta}|S_n(z)|^2,
\end{eqnarray}
and $|S_n (z)|^2 = S_n (z) \tilde{S}_n(z)$.

According to the definition in \cite{pezeshki2008doppler}, $G(z,\theta)$ is called a $z$-transform of the ambiguity function:
\begin{eqnarray}
g(k,\theta) = \sum_{n=0}^{N-1} e^{jn\theta}c_n (k),
\end{eqnarray}
where $c_n (k) = \sum_{l=0}^{L-1} s_n [l] s^*_n [l+k]$ is the auto-correlation function (ACF) of $\mathbf s_n$.

We hope that for any $\theta$ among a modest Doppler shift interval, $G(z,\theta)$ has very low range sidelobes in a proper range interval $[-Z_{max}, Z_{max}]$,  $Z_{max}\leq L$ . In other words, the desired  $G(z,\theta)$ should be
\begin{eqnarray}\label{eq: ghope}
G(z,\theta)= \alpha (\theta)+      \sum_{l=-Z_{max}+1\atop l\neq 0}^{Z_{max}-1} v_l z^l + \hat {G}(z),
\end{eqnarray}
where $|v_l|\leq \delta$ for $l=-Z_{max}+1,\cdots,-1,1,\cdots,Z_{max}-1$, and $\delta$ is a very small positive real number. Also, $\hat {G}(z)$ is given by
\begin{equation}
\hat {G}(z)=\sum_{l=-L+1}^{-Z_{max}+1}v_l z^l + \sum_{l=Z_{max}-1}^{L-1}v_l z^l,
\end{equation}
which is what we do not pay attention on, because the range is outside the range interval $[-Z_{max}, Z_{max}]$. Besides, if $L=Z_{max}$, $\hat {G}(z)$ will vanish.

Now, we consider what is the key ingredient that can eliminate the Doppler effect to achieve the formula (\ref{eq: ghope}). Indeed, Taylor expansion used in \cite{pezeshki2008doppler} is a very important tool that can transform the wish into finding a solution to make the first $M$ Taylor coefficients almost vanish at all  desired nonzero delays.

The Taylor expansion of $G(z,\theta)$  around $\theta=0$ is  given by
\begin{eqnarray}
G(z,\theta) = \sum_{m=0}^{\infty} C_m(z)(j\theta)^m, \label{eq:taylor}
\end{eqnarray}
where $C_m (z)$ is the $m$-th Taylor coefficient which is given by
\begin{eqnarray}
C_m(z) = \sum_{n=0}^{N-1}n^m |S_{n}(z)|^2 \label{eq:taylorc}
\end{eqnarray}
for $ m=0,1,2,\cdots$.

In order to make $C_0(z), C_1(z),\cdots, C_M(z)$ almost vanish at all desired non-zero delays, $S_n(z)$ should be chosen carefully.
Note that, here PTM sequence is used to determine the sequence formed using  PTM sequence is still used to determine the sequence formed using ($\mathbf x$, $\mathbf y$). Therefore, in the $n$-th PRI, $S_n(z)$ is given by
\begin{eqnarray}
S_n (z)=(1-a_n)X(z)+a_n Y(z) \label{eq:PTMSn}
\end{eqnarray}
where $ \{a_n\}_{n=0}^{N-1}$ is the PTM sequence defined as the following recursions:
\begin{equation}
	\begin{cases}
	a_0=0;\\
	a_{2k}=a_k;\\
	a_{2k+1}=1-a_k;
	\end{cases}
\end{equation}
for all $k>0$.

\begin{theorem}
 If $|X|^2+|Y|^2$ satisfies the following
 \begin{eqnarray}
 |X|^2+|Y|^2=2L+\hat{\zeta}(z)+\check{\zeta}(z),
 \end{eqnarray}
 then $C_0(z), C_1(z),\cdots, C_M(z)$  can be almost vanished at all desired nonzero delays.
 Here
 \begin{eqnarray}
 \zeta(z)= \sum_{l=-Z_{max}+1\atop l\neq 0}^{Z_{max}-1} \hat{v}_l z^l,
 \end{eqnarray}
 and
 \begin{eqnarray}
\check{\zeta}(z)=\sum_{l=-L+1}^{Z_{max}-1}\check{v}_l z^l + \sum_{l=Z_{max}-1}^{L-1}\check{v}_l z^l,
 \end{eqnarray}
 where $\hat{v}_l\leq \hat{\delta}$, $\hat{\delta}<\delta$  and  $\check{v}_l$ is any value.
\end{theorem}

\begin{IEEEproof}
By substituting (\ref{eq:PTMSn}) into $C_m (z)$ shown in (\ref{eq:taylorc}), it is easy to verify
\begin{eqnarray}
\begin{split}
&C_m(z) =\sum_{n=0}^{N-1} n^m|(1-a_n)X+a_n Y|^2\\
&\quad=\sum_{n=0}^{N-1}n^m\left((1-a_n)^2|X|^2+a_n^2 |Y|^2\right)\\
&\quad\quad+\sum_{n=0}^{N-1} n^m\left((1-a_n)a_n^*XY^*+(1-a_n)^*a_n X^*Y \right)\\
&\quad=\left( \sum_{n=0}^{N-1}(1-a_n)n^m\right)|X|^2+\left(\sum_{n=0}^{N-1}a_n n^m\right)|Y|^2\\
&\quad=\beta_m (|X|^2+|Y|^2), \,\mathrm{for} \,\, m=0,1,2,\cdots,M,
 \end{split} \label{eq:cm2}
\end{eqnarray}
where   $\beta_m=\sum_{n=0}^{N-1} (1-a_n)n^m=\sum_{n=0}^{N-1}a_n n^m$  according to the Prouhet theorem (Theorem 1 in \cite{pezeshki2008doppler}), and $N=2^{M+1}-1$. It is noted that the third equation is satisfied based on the fact that the second part of the second equation can be vanished as $a_n$ can only be 0 or 1. Now, substituting $2L+\hat{\zeta}(z)+\check{\zeta}(z)$ into $C_m (z)$, then $C_m(z)$ is given by
\begin{eqnarray}
\begin{split}
& C_m (z)  = \beta_m (|X|^2+|Y|^2)= \beta_m (2L+\hat{\zeta}(z)+\check{\zeta}(z))= 2L\beta_m+\hat{\zeta}(z)\beta_m+\check{\zeta}(z)\beta_m.
\end{split}
\end{eqnarray}
The desired nonzero delays of $C_m (z)$ is $\hat{\zeta}(z)\beta_m$, which is written as
\begin{eqnarray}
\hat{\zeta}(z)\beta_m = \sum_{l=-Z_{max}+1\atop l\neq 0}^{Z_{max}-1} \beta_m \hat{v}_l z^l,
\end{eqnarray}
Since $\hat{v}_l$ is small enough, then  $\beta_m \hat{v}_l$ is still very small. Therefore, we can say that
$C_0(z)$, $C_1(z)$,$\cdots$, $C_M(z)$  can be almost vanished at all desired nonzero delays.
\end{IEEEproof}

\begin{remark}
The definition of quasi-Z-complementary pair $(\mathbf{x},\mathbf{y})$ in $z$-field is
\begin{eqnarray}
 |X|^2+|Y|^2=2L+\hat{\zeta}(z)+\check{\zeta}(z),
 \end{eqnarray}
 which is shown in Theorem 1.
\end{remark}

\begin{remark}
The amplitudes of
$C_{M+1}(z),C_{M+2}(z),\cdots, C_{\infty}(z)$ are not almost equal to zero at every desired nonzero delay, since
\begin{eqnarray}
	\sum_{n=0}^{N-1} (1-a_n)n^m \neq \sum_{n=0}^{N-1}a_n n^m,
\end{eqnarray}
when $m>M$. However, $\theta$ is small enough to make $ \sum_{m=M+1}^{\infty} C_m(z) (j\theta)^m$ convergent.
\end{remark}

\subsection{Application of QOZCPs in fully polarimetric radar}
Fully polarimetric radar systems are equipped with vertically/horizontally (V/H) dual-dipole elements at every antenna to make sure the simultaneous occurrence of transmitting and receiving on two orthogonal polarizations\cite{calderbank2006instantaneous}\cite{pezeshki2008doppler}\cite{cui2017broadband}. Sequences  $\mathbf s_V=\{\mathbf s_{V,n} \}_{n=0}^{N-1}$ from vertically polarization and $\mathbf s_H=\{\mathbf s_{H,n}\}_{n=0}^{N-1}$ from horizontally polarization are transmitted together over $N$ pulse repetition intervals (PRIs). It is noted that the two sequence $\mathbf s_V$ and $\mathbf s_H$ are  formed from the designed sequence pair $\mathbf x,\mathbf y$ which will be obtained in this paper. In other words, $\mathbf s_{V,n} $ and $\mathbf s_{H, n}$ can be chosen from $\left\{\pm\mathbf x, \pm\mathbf y, \pm \tilde{\mathbf x}, \pm\tilde{\mathbf y}\right\}$, where $\widetilde{\cdot}$ denotes reversed complex conjugate.  The $z$-transform of transmit matrix formed from $\mathbf s_V$ and $\mathbf s_H$ can be denoted as
\begin{equation}
\begin{split}
\mathbf{S}(z)
&=\left[\begin{matrix}
\mathbf S_V (z)\\
\mathbf S_H(z)
\end{matrix}\right]=\left[\begin{matrix}
S_{V,0} (z)&S_{V,1} (z)&\cdots&S_{V,N-1} (z)\\
S_{H,0} (z)&S_{H,1} (z)&\cdots&S_{H,N-1} (z)
\end{matrix}\right].
\end{split}
\end{equation}
The received matrix can be written as
\begin{eqnarray}
\mathbf{R}(z) = z^{-d_0}\mathbf{H S}(z)\mathbf{D}(\theta_0)+\mathbf{W}(z).
\end{eqnarray}
where $\mathbf{H}$ is the scattering matrix given by (\ref{eq:sm}), $\mathbf{W}(z)$ is a noise matrix, and $\mathbf{D}(\theta)$ is the Doppler modulation matrix which can be written as $\mathbf{D}(\theta) = \mathrm{diag}(1,e^{j\theta},\cdots,e^{j(N-1)\theta})$.

If the received matrix $\mathbf{R}(z)$ is processed by a filter matrix $\mathbf{\widetilde{S}}(z)$ written as
\begin{eqnarray}
\mathbf{\widetilde{S}}(z)=
\left[\begin{matrix}
\widetilde S_{V,0} (z)& \widetilde S_{V,1} (z)&\cdots&\widetilde S_{V,N-1} (z)\\
\widetilde S_{H,0} (z)&\widetilde S_{H,1} (z)&\cdots&\widetilde S_{H,N-1} (z)
\end{matrix}\right]^T,
\end{eqnarray}
then the receiver output is
\begin{eqnarray}
\mathbf{R}(z) \mathbf{\widetilde{S}}(z)= z^{-d_0}\mathbf{H}\mathbf{G}(z,\theta_0)+\mathbf{W}(z)\widetilde{\mathbf{S}}(z),
\end{eqnarray}
where $d_0$ is the delay coordinate of a point target.
The matrix $\mathbf{G}(z,\theta_0) = \mathbf{S}(z)\mathbf{D}(\theta_0)\widetilde{\mathbf{S}}(z)$ is defined as $z$-transform of a matrix-valued ambiguity function for $\mathbf{S}(z)$.
The scattering matrix $\mathbf{H}$ can be easily obtained on a pulse-by-pulse basis\cite{pezeshki2008doppler}, if $\mathbf{G}(z,\theta_0)$
has the following expression:
\begin{equation}
\begin{split}
&\mathbf{G}(z,\theta) =  \mathbf{S}(z)\mathbf{D}(\theta)\widetilde{\mathbf{S}}(z)\\
&=\left[\begin{matrix}
G_{VV}(z,\theta) & G_{VH}(z,\theta)\\
G_{HV}(z,\theta) & G_{HH}(z,\theta)
\end{matrix}\right]\approx \alpha(\theta)\left[\begin{matrix}
NL&0\\
0&NL
\end{matrix}\right],\label{eq: purpose}
\end{split}
\end{equation}
where $\alpha(\theta)$ is a function of $\theta$ independent of delay. Because $G_{VV}$ is equivalent to $G_{HH}$, and $G_{VH}=\widetilde{G}_{HV}$, it is sufficient to analyze $G_{VV}$ and $G_{VH}$ which are given by
\begin{eqnarray}
\begin{split}
&G_{VV} (z,\theta)= \sum_{n=0}^{N-1} e^{jn\theta} |S_{V,n}(z)|^2,\\
&G_{VH} (z,\theta)= \sum_{n=0}^{N-1} e^{jn\theta} S_{V,n}(z) \widetilde{S}_{H,n}(z),
\end{split}
\end{eqnarray}
where $|S_{V,n}(z)|^2$ is a $z$-transform of auto-correlation function of $s_{V,n}$ and $S_{V,n}(z) \widetilde{S}_{H,n}(z)$ is a $z$-transform of cross-correlation function of $s_{V,n}$ and $s_{H,n}$. The auto-correlation function of $s_{V,n}$ and  cross-correlation function of $s_{V,n}$ and $s_{H,n}$ are respectively given by
$C_{s_{V,n}} (k) = \sum_{l=0}^{L-1} \mathbf s_{V,n}(l) \mathbf s_{V,n}(l+k) ^*$ and $C_{s_{V,n},s_{H,n}} (k) = \sum_{l=0}^{L-1} \mathbf s_{V,n}(l) \mathbf s_{H,n}(l+k) ^*$, $l=0,1,\cdots,L-1$, where * denotes conjugate.

Again, we consider what is the key ingredient that can eliminate the Doppler effect to achieve the formula (\ref{eq: purpose}).  Taylor expansion is again used to transform the wish into finding a solution to make the first $M$ Taylor coefficients vanish at all desired nonzero delays.

The Taylor expansions of $G_{VV}(z,\theta)$ and $G_{VH}(z,\theta)$ around $\theta=0$ are respectively given by
\begin{eqnarray}
G_{VV}(z,\theta) = \sum_{m=0}^{\infty} CV_m(z)(j\theta)^m,G_{VH}(z,\theta) = \sum_{m=0}^{\infty} B_m(z)(j\theta)^m,
\end{eqnarray}
where
\begin{eqnarray}
CV_m(z) = \sum_{n=0}^{N-1}n^m |S_{V,n}(z)|^2,\\ \quad B_m(z) = \sum_{n=0}^{N-1}n^m S_{V,n}(z)\widetilde{S}_{H,n}(z), \label{eq:cm}
\end{eqnarray}
for $ m=0,1,2,\cdots$.

Combining the PTM sequence with the Alamouti matrix, we define a PTM-A matrix as follows
\begin{eqnarray}
\begin{split}
&\left[\begin{matrix}
S_{V,2k} & S_{V,2k+1}\\
S_{H,2k} & S_{H,2k+1}
\end{matrix}\right]=\left[\begin{matrix}
(1\!-\!a_{2k})X \!+\! a_{2k} (\!-\widetilde{Y}) & a_{2k+1}(\!-\! \widetilde{Y})\! \!+\!\! (1\!-\! a_{2k+1}) (\!-\!X)\\
\!\!(1-a_{2k})Y \!+\! a_{2k} \widetilde{X}  & a_{2k+1}\widetilde{X} \! +\!(1-a_{2k+1}) (-Y)
\end{matrix}\right],
\end{split}\label{eq:PTM-A}
\end{eqnarray}
where $ \{a_n\}_{n=0}^{N-1}$ is the PTM sequence defined above.

By substituting (\ref{eq:PTM-A}) into $CV_m (z)$ shown in (\ref{eq:cm}), and following the proof of theorem 1, it is easy to verify
\begin{eqnarray}
\begin{split}
&CV_m(z) = \beta_m (|X|^2+|Y|^2), \,\mathrm{for} \,\, m=0,1,2,\cdots,M,
 \end{split} \label{eq:cm2}
\end{eqnarray}
where   $\beta_m=\sum_{n=0}^{N-1} (1-a_n)n^m=\sum_{n=0}^{N-1}a_n n^m$  according to the Prouhet theorem (Theorem 1 in \cite{pezeshki2008doppler}), and $N=2^{M+1}-1$. Indeed, $CV_m (z)$ is definitely equivalent to $C_m(z)$.

Similarly, $B_m(z)$ can be written as
\begin{eqnarray}
\begin{split}
&B_m(z) \!\!= \!\!\!\!\sum_{k=0}^{N/2-1} \!\!(\!(1\!-\!2a_{2k})(2k)^m \!+\!(1\!-\!2a_{2k\!+\!1})(2k\!+\!1)^m \!)X\widetilde{Y}=\left[\sum_{n=0}^{N-1}(-1)^{a_{n}} n^{m}\right] X \widetilde{Y}.
\end{split}\label{eq:bm2}
\end{eqnarray}
Again, by Prouhet theorem, we can easily obtain
\begin{eqnarray}
\sum_{n=0}^{N-1}(-1)^{a_{n}} n^{m}=0, \, \mathrm{for}\, \,m=0,1,\cdots, M\label{eq:pt}.
\end{eqnarray}

The Taylor coefficients $CV_m$ and $B_m$ are clearly shown in (\ref{eq:cm2}) and (\ref{eq:bm2}), respectively.
From (\ref{eq:pt}), we know that $B_{0},B_{1},\cdots,B_{M}$ can vanish thoroughly for any $\mathbf x, \mathbf y$, which attributes to a PTM-A matrix. However, the $B_{M+1},B_{M+2},\cdots$ can not be equal to zero when $N=2^{M+1}-1$. Therefore, for eliminating all $\{B_m\}_{m=0}^{\infty}$, the $X\widetilde{Y}$ should be equal to zero. But one can hardly achieve the purpose. To compromise on it, the proposed QOZCP can help $X\widetilde{Y}$  meet the following low local cross-correlation demand:
\begin{eqnarray}
X\widetilde{Y} = \sum_{l=-L+1}^{L+1} u_l z^l, \label{eq:qo}
\end{eqnarray}
where $|u_l|\leq \delta$ for $l=-Z_{max}+1,\cdots,-1,0,1,\cdots,Z_{max}-1$, $Z_{max}\leq L$, and $\delta$ is very close to zero.
From (\ref{eq:cm2}), in order to make $CV_m=\beta_m$ with $m = 0,1,2,\cdots,M$, $|X|^2+|Y|^2$ should be equal to $2L$, which is exactly the property of Golay complementary pair. However, all GCPs cannot meet the low local cross-correlation demand, i.e., (\ref{eq:qo}). To avoid the problem, we want $X\widetilde{Y}$ to meet the following property of low local sum of autocorrelations:
\begin{eqnarray}
|X|^2+|Y|^2= 2L+  \sum_{l=-L+1\atop l\neq 0}^{L+1} v_l z^l\label{eq:qzc}
\end{eqnarray}
where $|v_l|\leq \delta$ for $l=\pm 1,\pm 2, \cdots, \pm (Z_{max}-1)$, $Z_{max}\leq L$, and $\delta$ is very close to zero.

\begin{theorem}
 If $X, Y$ satisfy the following equations
 \begin{eqnarray}
 |X|^2+|Y|^2=2L+\hat{\zeta}(z)+\check{\zeta}(z),X\widetilde{Y} = \xi(z)+\hat{\xi}(z),
 \end{eqnarray}

 then $CV_0(z), CV_1(z),\cdots, CV_M(z)$ tends to zero at all desired nonzero delays.
 Here
 \begin{eqnarray}
 \zeta(z)= \sum_{l=-Z_{max}+1\atop l\neq 0}^{Z_{max}-1} \hat{v}_l z^l,
 \xi(z) = \sum_{l=-Z_{max}+1}^{Z_{max}-1} \hat{u}_l z^l,
 \end{eqnarray}
 and
 \begin{eqnarray}
\check{\zeta}(z)=\sum_{l=-L+1}^{Z_{max}-1}\check{v}_l z^l + \sum_{l=Z_{max}-1}^{L-1}\check{v}_l z^l,
\check{\xi}(z)=\sum_{l=-L+1}^{Z_{max}-1}\check{u}_l z^l + \sum_{l=Z_{max}-1}^{L-1}\check{u}_l z^l,
 \end{eqnarray}
 where $\hat{v}_l,\hat{u}_l\leq \hat{\delta}$, and $\hat{\delta}<\delta$  and  $\check{v}_l,\check{u}_l$ are any value.
\end{theorem}

\section{Numerical Experiments and Discussions}

In this section,  we aim at comparing the proposed Doppler resilient (DR) QOZCPs with DR-GCP sequences on the performance of Doppler resilience.  Besides, the complementary sidelobes and cross correlation of $(L,Z)$-QOZCPs ($\mathrm{PAPR}=5$) obtained by SDAMM are also compared with those of length-$L$ GCPs .
The two DR-QOZCP sequences and the two DR-GCP sequences are generated  by the $(L,Z)$-QOZCP $(\mathbf x_Q, \mathbf y_Q)$, and length-$L$ GCP $(\mathbf x_G, \mathbf y_G)$,  respectively. The PTM-A matrix is chosen as (\ref{eq:PTM-A}) with $N=8$. Then the transmit matrix for DR-QOZCP sequences and DR-GCP sequences is chosen as follows,
\begin{eqnarray}
\mathbf{s}
=\left[\begin{matrix}
\mathbf {s_V}\\
\mathbf {s_H}
\end{matrix}\right]
=\left[\begin{matrix}
\mathbf x&- \tilde{\mathbf y}&- \tilde{\mathbf y}&-\mathbf x&- \tilde{\mathbf {y}}&-\mathbf x&\mathbf x&- \tilde{\mathbf y}\\
\mathbf {y}&\tilde{\mathbf {x}}& \tilde{\mathbf {x}}&-\mathbf {y}& \tilde{\mathbf {x}}&-\mathbf {y}&\mathbf{y}& \tilde{\mathbf {x}}
\end{matrix}\right],\end{eqnarray}
where $\mathbf {s_V}$ and $\mathbf {s_H}$ are Doppler resilient sequences sent from V polarization direction and   from H polarization direction, respectively.
Their auto-ambiguity function (AAF) $g_{VV} (k,\theta)$ of $\mathbf s_V$, and cross-ambiguity function (CAF) $g_{VH} (k,\theta)$ of $\mathbf s_V$, $\mathbf s_H$ are respectively given by
\begin{eqnarray}
\begin{split}
&g_{VV} (k,\theta)= \sum_{n=0}^{N-1} e^{jn\theta} C_{VV,n}(k)=\sum_{n_0={0,3,5,7}}e^{jn_0\theta}C_{\mathbf x} (k)+\sum_{n_1={1,2,4,6}}e^{jn_1\theta}C_{\mathbf y} (k),
\end{split}\label{eq:gvv}
\end{eqnarray}
and
\begin{eqnarray}
\begin{split}
&g_{VH} (k,\theta)= \sum_{n=0}^{N-1} e^{jn\theta} C_{VH,n}(k)=\sum_{n_0={0,3,5,7}}e^{jn_0\theta}C_{xy} (k)-\sum_{n_1={1,2,4,6}}e^{jn_1\theta}C_{xy} (k).
\end{split}\label{eq:gvh}
\end{eqnarray}

 \begin{table*}[]
\centering
\caption{\normalsize\\ \normalsize Maximun Complementary Sidelobe and Maximun Cross-Correlation of $(\mathbf x, \mathbf y)$ with $\mathrm{PAPR}=5$}
\label{tab:my-table}
\begin{tabular}{|l|l|r|l|l|l|}
\hline
\multicolumn{1}{|c|}{\multirow{2}{*}{L}} & \multicolumn{1}{c|}{\multirow{2}{*}{Z}} & \multicolumn{2}{r|}{Maximun Complementary Sidelobe in $[-Z+1,Z-1]$} & \multicolumn{2}{l|}{Maximun Cross-Correlation in $[-Z+1,Z-1]$} \\ \cline{3-6}
\multicolumn{1}{|c|}{}   & \multicolumn{1}{c|}{}   & \multicolumn{1}{c|}{\quad \quad\quad GCP \quad \quad\quad}   &  \multicolumn{1}{c|}{\quad \quad QOZCP \quad \quad} & \multicolumn{1}{c|}{\quad \quad \quad GCP \quad\quad \quad}    & \multicolumn{1}{c|}{\quad \quad QOZCP \quad \quad} \\ \hline
64  & 10   & \multicolumn{1}{c|}0  & \multicolumn{1}{c|}{ $6.10\times10^{-11}$}  &  \multicolumn{1}{c|}{14}  &  \multicolumn{1}{c|}{$1.90\times10^{-10}$} \\ \hline
64  & 15   & \multicolumn{1}{c|}0  & \multicolumn{1}{c|}{ $1.31\times10^{-10}$}  &  \multicolumn{1}{c|}{15}  &  \multicolumn{1}{c|}{$4.56\times10^{-10}$} \\ \hline
64  & 20   & \multicolumn{1}{c|}0  & \multicolumn{1}{c|}{ $3.99\times10^{-10}$}  &  \multicolumn{1}{c|}{15}  &  \multicolumn{1}{c|}{$1.09\times10^{-9}$} \\ \hline
64  & 25   & \multicolumn{1}{c|}0  &\multicolumn{1}{c|}{ $6.18\times 10^{-10}$}  &  \multicolumn{1}{c|}{15}  &  \multicolumn{1}{c|}{$1.71\times10^{-9}$} \\ \hline
64  & 30   & \multicolumn{1}{c|}0  &\multicolumn{1}{c|}{ $9.46\times 10^{-10}$}                        &  \multicolumn{1}{c|}{15}  &  \multicolumn{1}{c|}{$2.09\times 10^{-9}$} \\ \hline
\end{tabular}
\end{table*}

\begin{table*}[]
\centering
\caption{\\ \normalsize Maximun Auto-Ambiguity Function Sidelobes and Maximun Cross-Ambiguity Functions of $(\mathbf x, \mathbf y)$}
\label{tab:my-table}
\begin{tabular}{|l|l|r|l|l|l|}
\hline
\multicolumn{1}{|c|}{\multirow{2}{*}{L}} & \multicolumn{1}{c|}{\multirow{2}{*}{Z}} & \multicolumn{2}{r|}{Maximun Auto-Ambiguity Function Sidelobes in $\Omega_1$} & \multicolumn{2}{l|}{Maximun Cross-Ambiguity Function in $\Omega_2$} \\ \cline{3-6}
\multicolumn{1}{|c|}{}   & \multicolumn{1}{c|}{}   & \multicolumn{1}{c|}{\quad \quad\quad DR-GCP \quad \quad\quad}   &  \multicolumn{1}{c|}{\quad \quad DR-QOZCP \quad \quad} & \multicolumn{1}{c|}{\quad \quad \quad DR-GCP \quad \quad \quad}    & \multicolumn{1}{c|}{\quad \quad DR-QOZCP \quad \quad} \\ \hline
64  & 10   & \multicolumn{1}{c|}{$5.39\times 10^{-2}$}  & \multicolumn{1}{c|}{ $5.52\times10^{-2}$}  &  \multicolumn{1}{c|}{76.98}  &  \multicolumn{1}{c|}{$1.04\times10^{-9}$}   \\ \hline
64  & 15   & \multicolumn{1}{c|}{$6.93\times 10^{-2}$}  & \multicolumn{1}{c|}{ $6.76\times10^{-2}$}  &  \multicolumn{1}{c|}{76.98}  &  \multicolumn{1}{c|}{$2.51\times10^{-9}$}    \\ \hline
64  & 20   & \multicolumn{1}{c|}{$6.93\times 10^{-2}$}  & \multicolumn{1}{c|}{ $7.66\times10^{-2}$}  &  \multicolumn{1}{c|}{82.48}  &  \multicolumn{1}{c|}{$5.98\times 10^{-9}$} \\ \hline
64  & 25   & \multicolumn{1}{c|}{$1.00\times 10^{-1}$}  &\multicolumn{1}{c|}{ $8.53\times 10^{-2}$}                        &  \multicolumn{1}{c|}{82.48}  &  \multicolumn{1}{c|}{$9.40\times 10^{-9}$} \\ \hline
64  & 30   & \multicolumn{1}{c|}{$1.00\times 10^{-1}$}  &\multicolumn{1}{c|}{ $8.61\times 10^{-2}$}                        &  \multicolumn{1}{c|}{$82.48$}  &  \multicolumn{1}{c|}{$1.15\times 10^{-8}$} \\ \hline
\end{tabular}
\end{table*}

\begin{figure} \centering
\subfigure[The sum of auto-correlation functions, i.e., $|C_x(k)+C_y(k)|$.] {
 \label{fig:a}
\includegraphics[width=0.45\columnwidth]{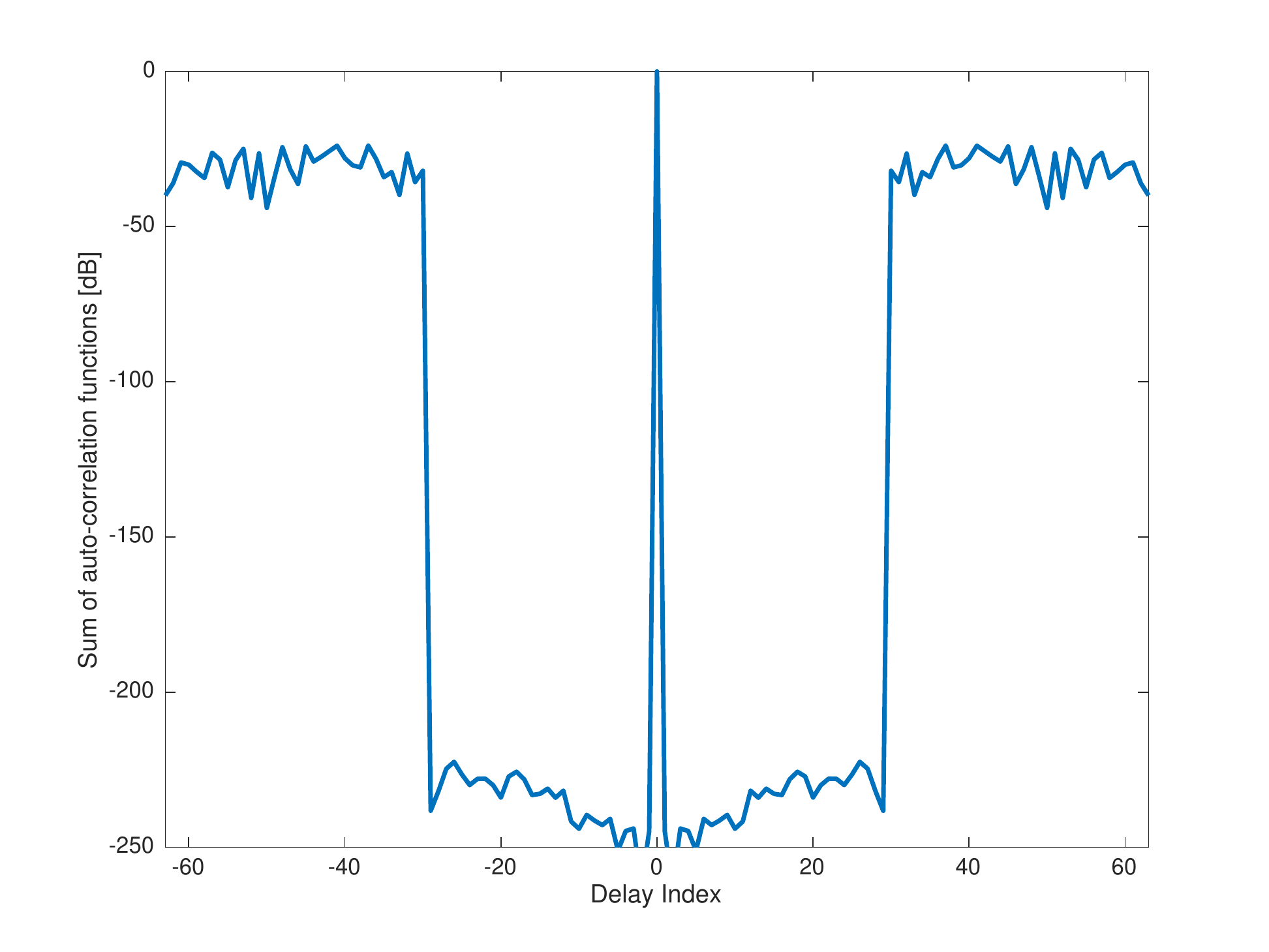}
}
\subfigure[The cross-correlation function, i.e., $|C_{xy}|(k)$] {
\label{fig:b}
\includegraphics[width=0.45\columnwidth]{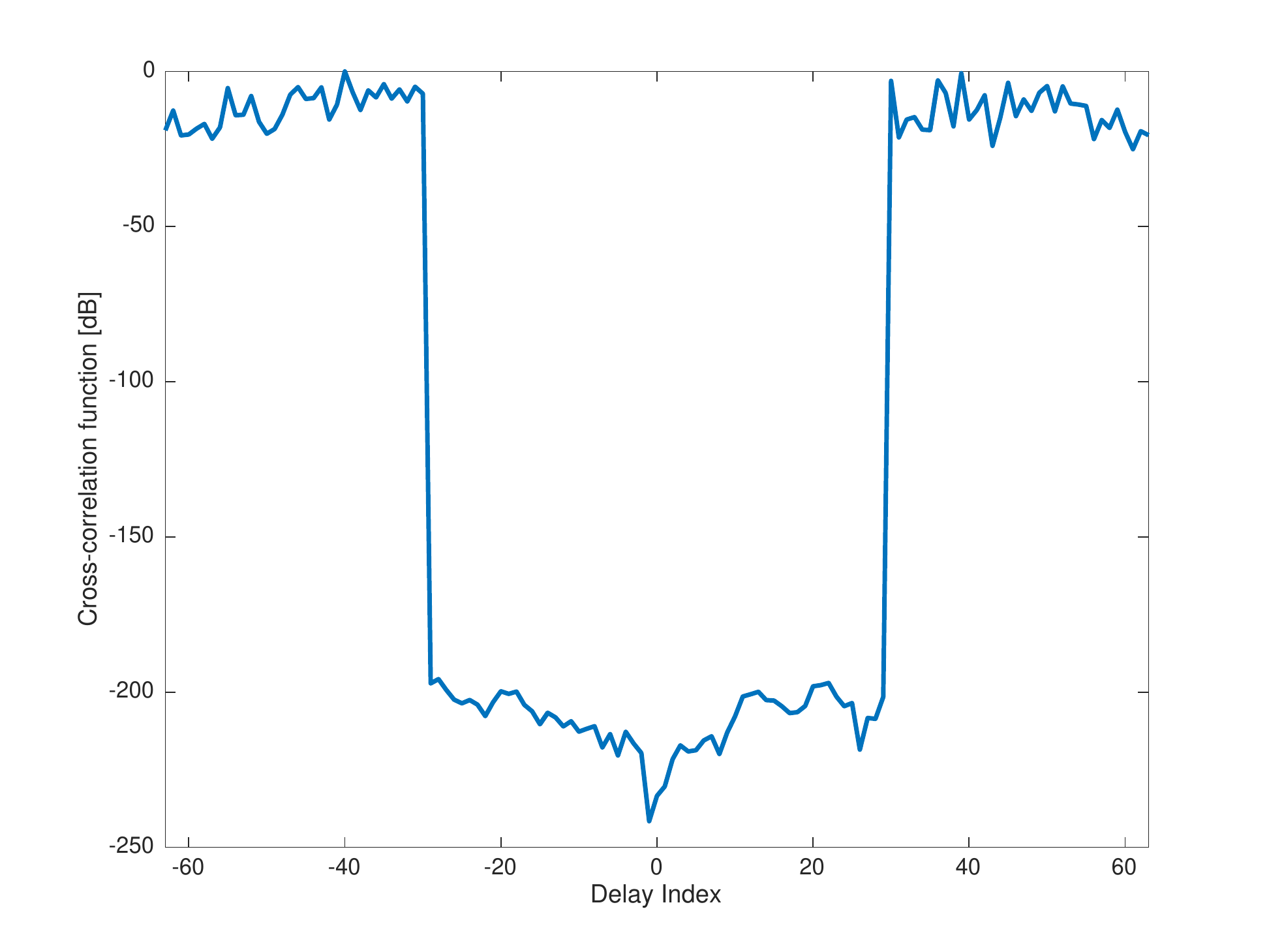}
}
\caption{ Two figures about the QOZCP $(\mathbf x,\mathbf y)$ by the Algorithm 1 }
\label{fig:acf}
\end{figure}

\begin{figure} \centering
\subfigure[The AAF of a DR-GCP sequence.] {
 \label{fig:a}
\includegraphics[width=0.475\columnwidth]{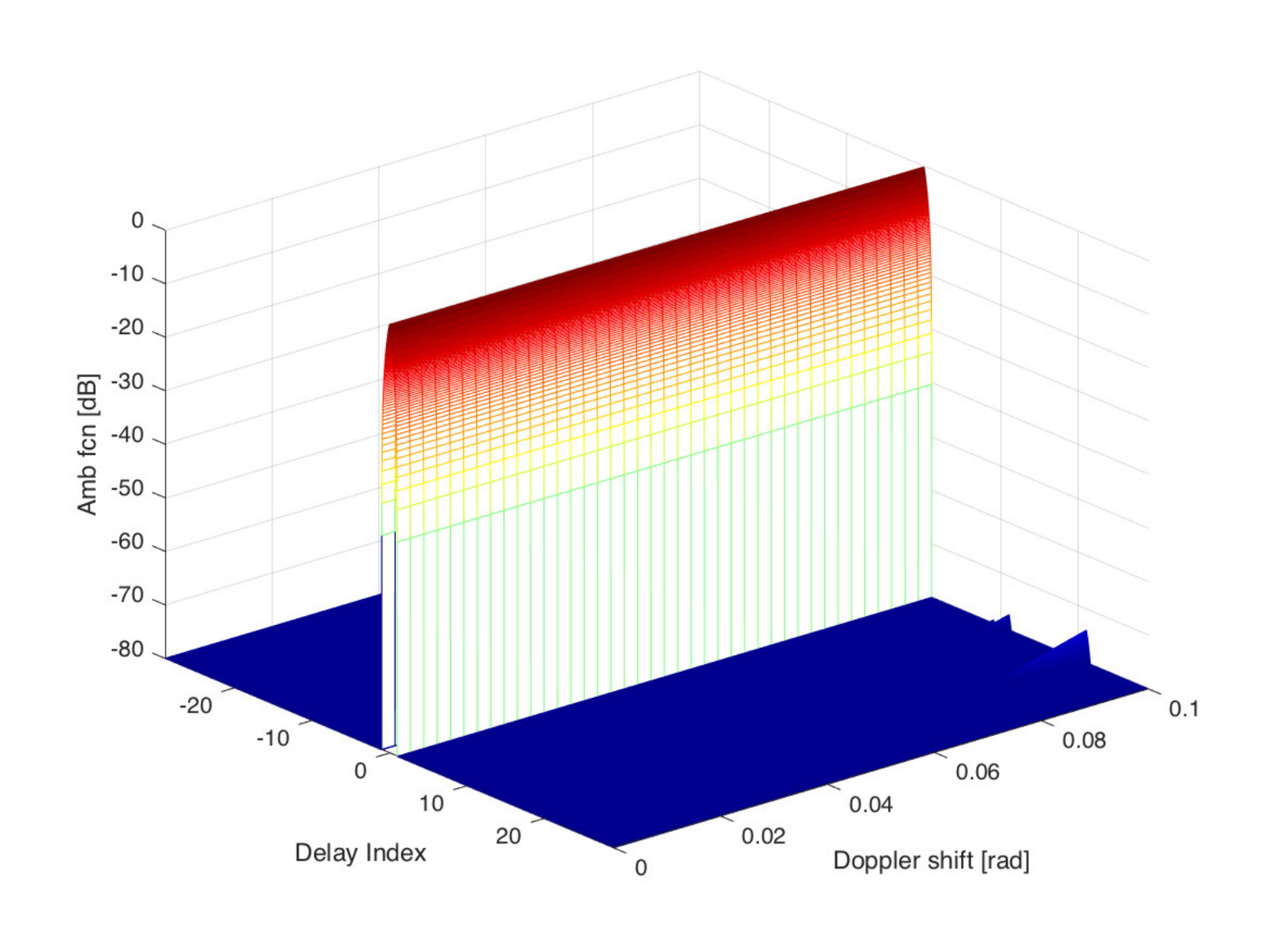}
}
\subfigure[The AAF of the a DR-QOZCP sequence] {
\label{fig:b}
\includegraphics[width=0.475\columnwidth]{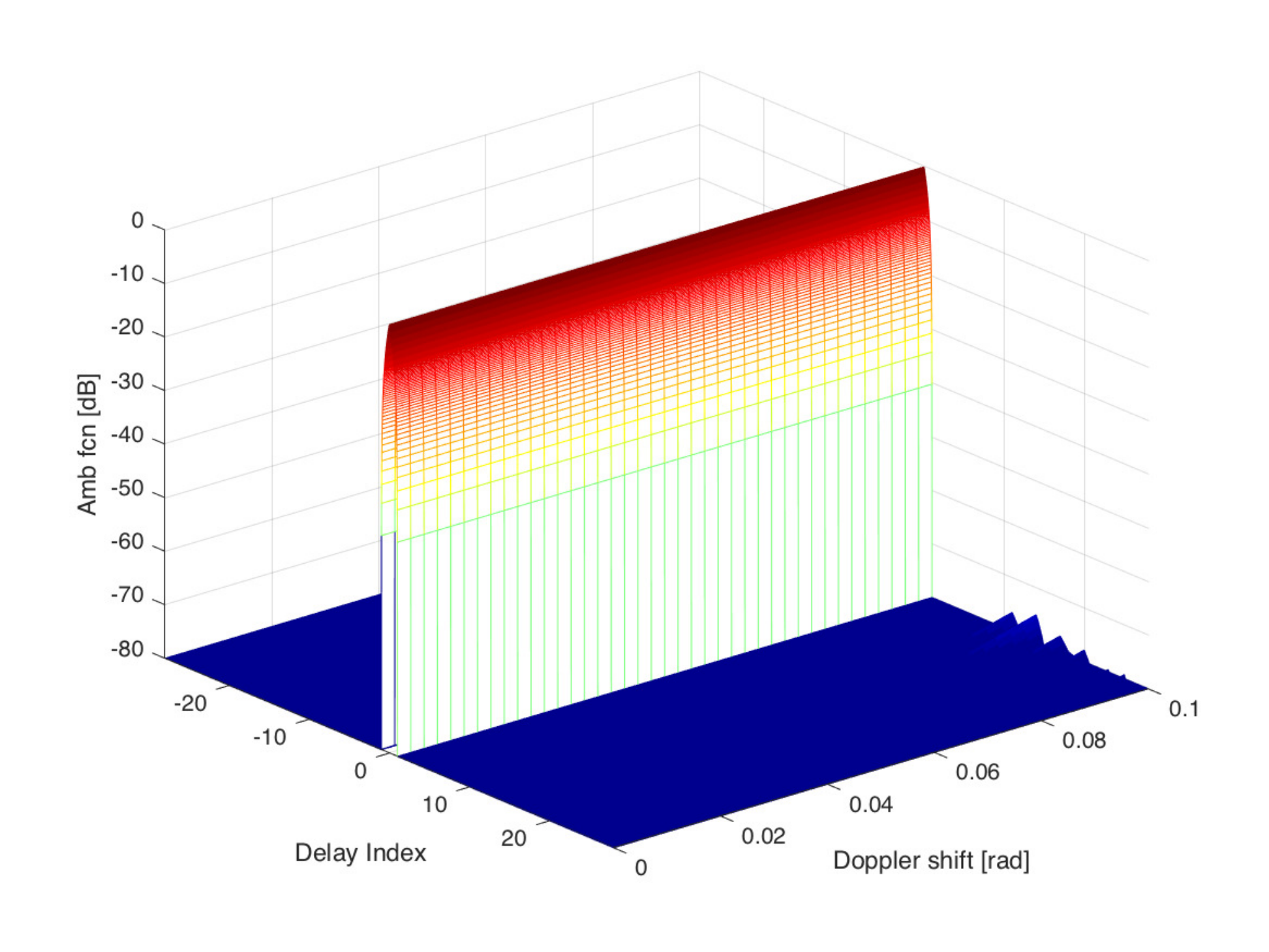}
}
\caption{  AAFs of the DR-GCP sequence and the DR-QOZCP sequence, respectively}
\label{fig:amb}
\end{figure}

Remark: For computing their AAF and CAF of DR-QOZCP sequence(s), $\mathbf x_Q$, $\mathbf y_Q$, and  $\mathbf x_G$, $\mathbf y_G$ can be substituted into $\mathbf x$, $\mathbf y$  in (\ref{eq:gvv}) and (\ref{eq:gvh}).

\begin{figure}[htbp]
\centering
\subfigure[The CAF of two DR-GCP sequences]{
\begin{minipage}{7cm}
\centering
\includegraphics[width=8cm]{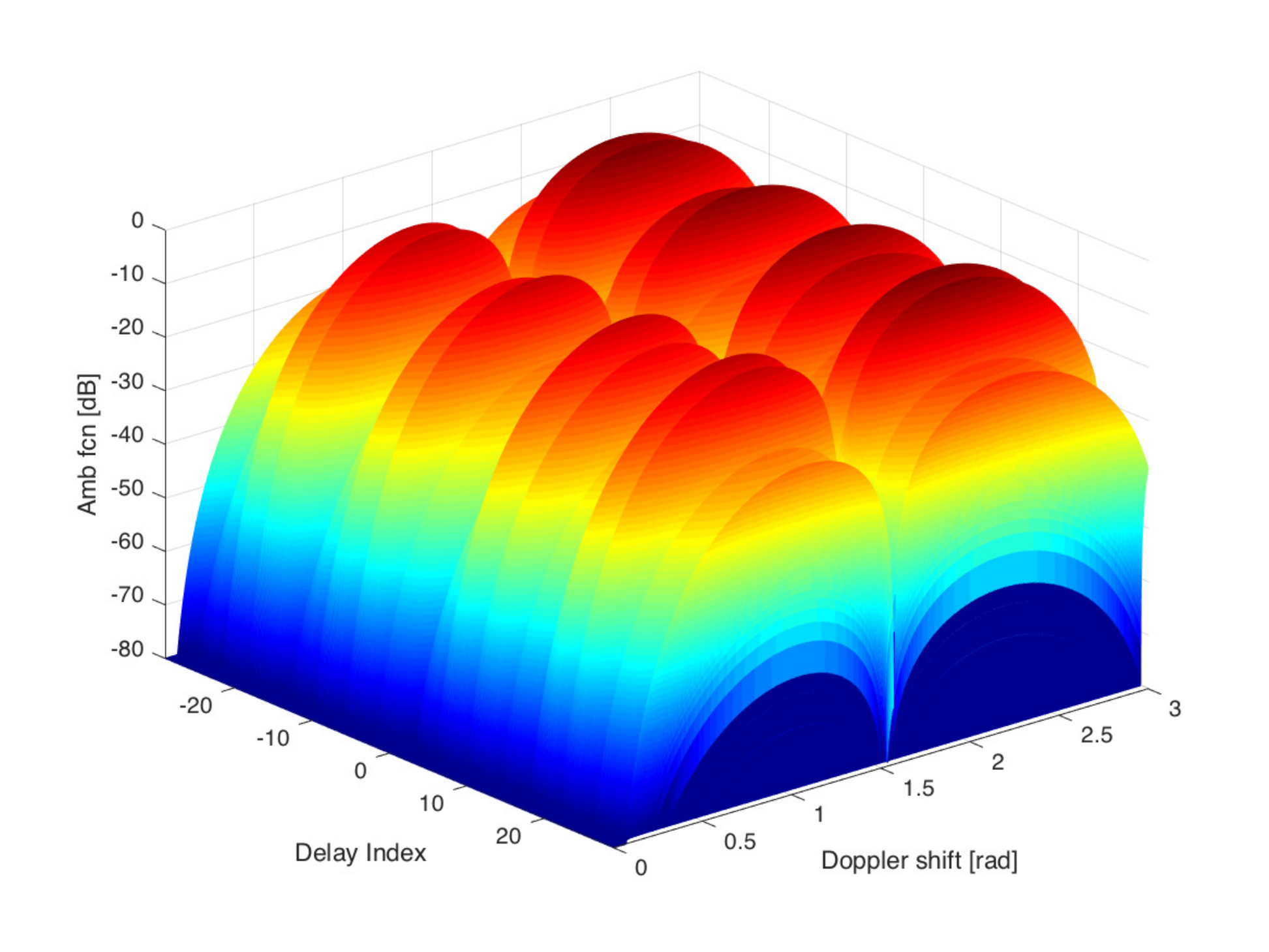}\\
\end{minipage}
}
\subfigure[The CAF of two DR-QOZCP sequences]{
\begin{minipage}{8cm}
\centering
\includegraphics[width=8cm]{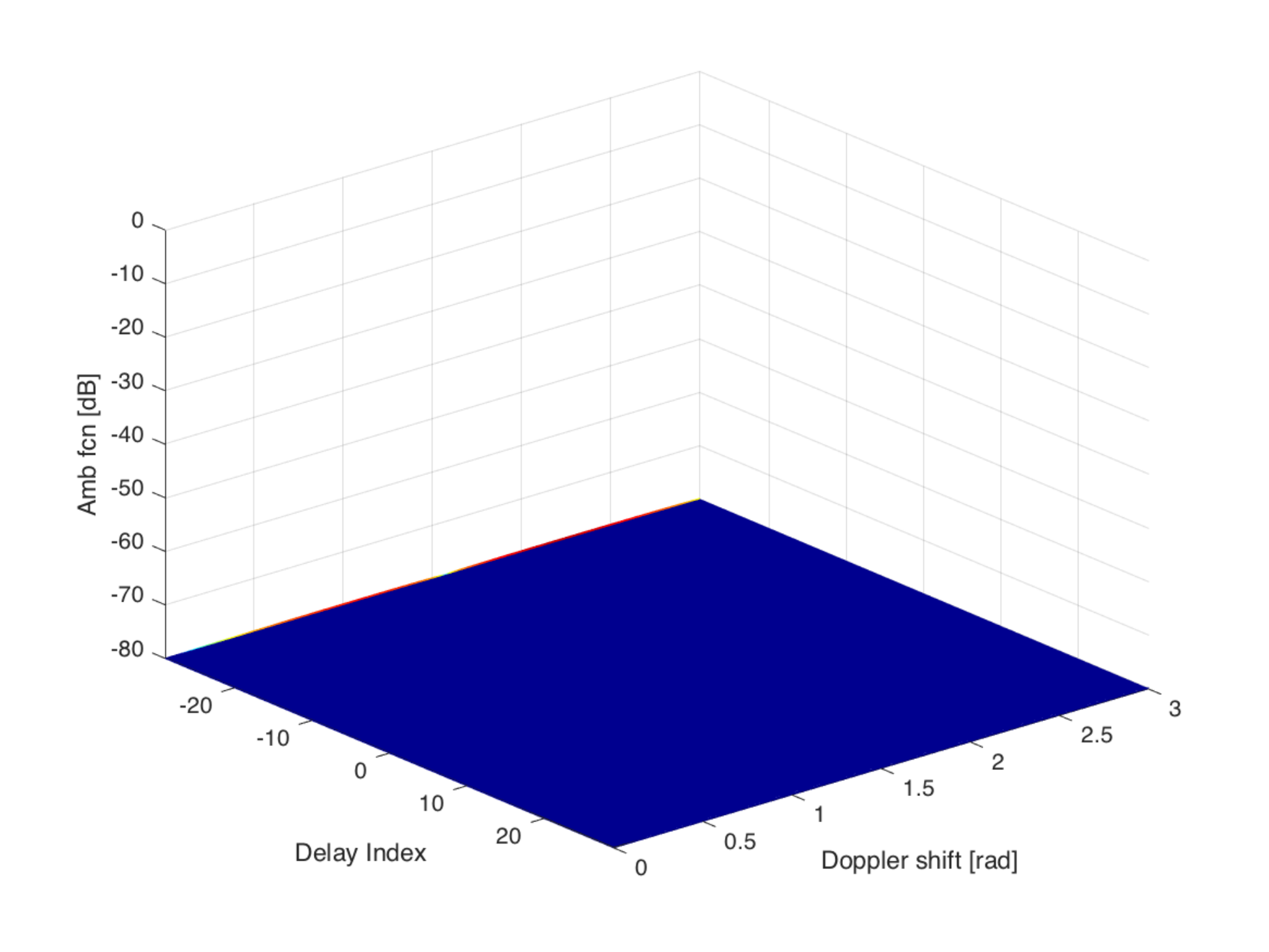}
\end{minipage}
}
\caption{ CAFs of DR-GCP sequences and DR-QOZCP sequences, respectively}
\label{fig:camb}
\end{figure}

Some conclusions can be drawn based on Table I, Table II, and  Fig. \ref{fig:acf} to Fig. \ref{fig:camb}.
\begin{itemize}
\item At first, we will compare $(L,Z)$-QOZCP ($\mathrm{PAPR}=5$) with length-$L$ GCP on maximun complementary sidelobe and cross correlation in delay interval $[-Z+1,Z-1]$.  As Table I shows, there is a significant difference between QOZCP and GCP. Although QOZCP and GCP both have good complementary sidelobe in delay interval $[-Z+1,Z-1]$, QOZCP has much lower cross correlation than GCP in $[-Z+1,Z-1]$.

\item As shown in Fig. \ref{fig:acf}, two figures about the obtained $(L,Z)$-QOZCP $(\mathbf x,\mathbf y)$ by  SDAMM, where $L=64$,$Z=30$, $\mathrm{PAPR}=5$. Fig. \ref{fig:acf}(a) is the sum of correlation functions $C_x (k)+C_y (k)$ and  Fig. \ref{fig:acf}(b) is the cross-correlation function $C_{xy}(k)$. The sidelobe in Fig. \ref{fig:acf}(a) is very low in  delay interval $[-Z+1,Z-1]$. The cross-correlation level in  interval $[-Z+1,Z-1]$ in Fig. \ref{fig:acf}(b) is also very low.

\item Besides,  we will compare DR-QOZCP with DR-GCP on Maximun Auto-Ambiguity Function Sidelobes in $\Omega_1$ and Maximun Cross-Ambiguity Function in $\Omega_2$ , where $\Omega_1 = \{(k,\theta)| k\in [-Z+1,Z-1], \theta \in [0,0.1]\}$ and $\Omega_2 = \{(k,\theta)| k\in [-Z+1,Z-1], \theta \in [0,3]\}$. What is interesting in table II is that DR-QOZCPs derived by $(L,Z)$-QOZCP have  much lower Maximun Cross-Ambiguity Function than DR-GCPs', while Maximun Auto-Ambiguity Function Sidelobes are low for both DR-GCP and DR-QOZCP.

\item  Fig. \ref{fig:amb}(a) and Fig. \ref{fig:amb}(b) show an AAF of the DR-GCP sequence and that of the DR-QOZCP sequence in the Delay interval $[-Z+1, Z-1]$ and Doppler interval $[0, 0.1]$. The sidelobes of the designed sequence in delay interval $[-Z+1,Z-1]$ are almost as good as that of the DR-GCP sequence along any Doppler shift in $[0, 0.1]$.

\item Fig. \ref{fig:camb}(a) and Fig. \ref{fig:camb}(b) show an AAF of a DR-GCP sequence and that of the designed DR-QOZCP sequence, respectively, in the Delay interval $[-20,20]$ and Doppler interval $[0, 3]$. The level of Fig. \ref{fig:camb}(b) is significantly lower than that of Fig. \ref{fig:camb}(a).  For the system, the cross polarization aliasing is sufficiently vanished along any Doppler shift in $[0,3]$ when the proposed sequence pair is used.
\end{itemize}

In short, the Doppler resilient performance of our DR-QOZCP sequences is significantly better than that of DR-GCP sequences.

\section{Concluding Remarks and Open Problems}

In this paper we have given two  contributions. Firstly, we have proposed new sets of sequence pairs called ``Quasi-Orthogonal Z-Complementary Pairs" (QOZCPs). Secondly, we have applied the proposed QOZCPs to design Doppler resilient waveforms for polarimetric radar systems with are also dual orthogonal. We have also solved an optimization  problem efficiently by proposing SADMM algorithm in MM framework, which eventually constructs the proposed QOZCPs of any length. Finally, we have compared the ambiguity plots of the proposed QOZCPs with the ambiguity plots of the existing DR-GCPs. The comparison shows that the proposed QOZCPs are more efficient as compared to DR-GCPs when CAF plots are considered.

Owing to the efficiency of the QOZCPs in application scenarios, the reader is  invited to propose systematic constructions of QOZCPs and analyze the structural properties of these pairs.

\section*{Appendix A\\Proof of Proposition 1}
\begin{IEEEproof}
Calculate
\begin{eqnarray}
\begin{split}
\sum_{k=-L+1}^{L-1} w_k |\mathbf {z^H A_k z}|^2&=\sum_{k=-L+1}^{L-1} w_k \text{vec}(\mathbf Z)^H \text{vec}(\mathbf A_k) \text{vec}(\mathbf A_k)^H \text{vec}(\mathbf Z)\\
&= \text{vec}(\mathbf Z)^H \left[\sum_{k=-L+1}^{L-1} w_k \text{vec}(\mathbf A_k) \text{vec}(\mathbf A_k)^H \right]\text{vec}(\mathbf Z),
\end{split}
\end{eqnarray}
\begin{eqnarray}
\begin{split}
 &\sum_{k=-L+1}^{L-1} \tilde{w}_k  |\mathbf {z^H B_k z}|^2= \text{vec}(\mathbf Z)^H \left[\sum_{k=-L+1}^{L-1} \tilde{w}_k \text{vec}(\mathbf B_k) \text{vec}(\mathbf B_k)^H \right]\text{vec}(\mathbf Z).
 \end{split}
 \end{eqnarray}
We set $J_A = \sum_{k=-L+1}^{L-1} w_k \text{vec}(\mathbf A_k) \text{vec}(\mathbf A_k)^H$, and
 $J_B = \sum_{k=-L+1}^{L-1} w_k \text{vec}(\mathbf B_k) \text{vec}(\mathbf B_k)^H$.
Then, It is easy to verify that the objective of (\ref{opt:p00}) can be transformed into that of (\ref{opt:p01})
\end{IEEEproof}

\section*{Appendix B\\ Proof of proposition 2}
\begin{IEEEproof}
The objective of (\ref{opt:p01}) can be majorized by
\begin{eqnarray}
\begin{split}
&M_{f_1} (\mathbf Z, \mathbf Z^{(t)})\\
&=\lambda_{\mathbf J} \text{vec}(\mathbf Z)^H \text{vec}(\mathbf Z)+2 \text{Re}(\text{vec}(\mathbf Z)^H(\mathbf J-\lambda_{\mathbf J}\mathbf I)\text{vec}(Z^{(t)}))+\text{vec}(\mathbf Z^{(t)})^H(\lambda_{\mathbf J}\mathbf I-\mathbf J)vec(Z^{(t)})
\end{split}\label{eq:mf1}
\end{eqnarray}
The first term is  equal to a constant $\lambda_{\mathbf J} L^2$, and the last term is also a constant obviously. Therefore,  we ignore the both terms and only keep the second term, and the objective of (\ref{opt:p01}) can be majorized by the second term. It is noted that the coefficient 2 cannot make any effect on the optimization, so we remove it.
\end{IEEEproof}

\section*{Appendix C\\ Proof of Theorem 1}
\begin{IEEEproof}
Calculate
\begin{eqnarray}
J\text{vec}(\mathbf A_k)& =& (\alpha J_A+(1-\alpha)J_B) \text{vec}(\mathbf A_k)=\alpha J_A \text{vec}(\mathbf A_k)\\
&=&\sum_{i=1-L}^{L-1} w_i \alpha \text{vec}(A_i) \text{vec}(A_i)^H\text{vec}(A_k)\\
&=&w_k \alpha \text{vec}(A_k) \text{vec}(A_k)^H\text{vec}(A_k)\\
&=&w_k \alpha (2N-2|k|) \text{vec}(A_k).
\end{eqnarray}
Therefore,
\begin{eqnarray}
\lambda_{A}(k) = w_k \alpha (2N-2|k|).
\end{eqnarray}

 Calculate
 \begin{eqnarray}
J\text{vec}(\mathbf B_k) &=& (J_A+J_B) \text{vec}(\mathbf B_k)=J_B \text{vec}(\mathbf B_k)\\
&=&\sum_{i=1-L}^{L-1} w_i \alpha \text{vec}(B_i) \text{vec}(B_i)^H\text{vec}(B_k)\\
&=&w_k \alpha \text{vec}(B_k) \text{vec}(B_k)^H\text{vec}(B_k)\\
&=&w_k (1-\alpha) (N-|k|) \text{vec}(A_k).
\end{eqnarray}
Therefore,
 \begin{eqnarray}
\lambda_{B}(k) = w_k (1-\alpha) (N-|k|);
\end{eqnarray}

Because $\lambda_A (k)$ and $\lambda_B(k)$ are the eigenvalue of $\mathbf J$, then
\begin{eqnarray}
\lambda_{\mathbf J} = \max_k \left\{\max\{\lambda_{A}(k), \lambda_{B}(k) \right\}|k=-L+1,\cdots,L-1\}.
\end{eqnarray}
\end{IEEEproof}

\section*{Appendix D\\Proof of  Proposition 3}
\begin{IEEEproof}
The main operations are in  (\ref{eq:bigproof})
 \begin{figure*}
 \begin{eqnarray}
 \begin{split}
&\text{Re}\left\{\text{vec}(\mathbf Z)^H (\mathbf J-\lambda_{\text{max}}(\mathbf J)\mathbf I)\text{vec}(\mathbf Z^{(t)})\right\} \\
&= \text{Re}\left\{ \text{vec}(\mathbf Z)^H  \left( \alpha \sum_{k=-L+1}^{L-1} w_k \text{vec}(\mathbf A_k) \text{vec}(\mathbf A_k)^H \right.\right.\\
&\qquad\left.\left.+ (1-\alpha)\sum_{k=-L+1}^{L-1} \tilde{w}_k \text{vec}(\mathbf B_k) \text{vec}(\mathbf B_k)^H-\lambda_{\mathbf J}\mathbf I \right) \cdot\text{vec}(\mathbf Z^{(t)})     \right\}  \\
&= \text{Re}\left\{ \sum_{k=-L+1}^{L-1} \alpha w_k (\text{Tr}(\mathbf A_k \mathbf Z) \text{Tr}(\mathbf A_{-k} \mathbf Z^{(t)}))
\right.\\
&\left.\qquad+ \sum_{k=-L+1}^{L-1} (1-\alpha)\tilde{w}_k (\text{Tr}(\mathbf B_k \mathbf Z) \text{Tr}(\mathbf B_{-k} \mathbf Z^{(t)}))-\lambda_{\mathbf J} \text{Tr} (\mathbf Z^{(t)} \mathbf Z)\right\}\\
&= \text{Re}\left\{ \alpha \text{Tr} ( \sum_{k=-L+1}^{L-1} w_k r_{-k}^{(l)} \mathbf A_k \mathbf Z)
+(1-\alpha) \text{Tr} ( \sum_{k=-L+1}^{L-1} \tilde{w}_k c_{-k}^{(l)} \mathbf B_k \mathbf Z))-\lambda_{\mathbf J} \text{Tr}(\mathbf Z^{(t)}\mathbf Z)\right\} \\
&= \text{Re}\left\{ \alpha Tr(\mathbf Q_r \mathbf Z)+(1-\alpha) Tr(\mathbf Q_c \mathbf Z)-\lambda_{\text{max}}(\mathbf J) \text{Tr}(\mathbf Z^{(t)}\mathbf Z) \right\} \\
&= \text{Re}\left\{ \mathbf z^H \left(\alpha \mathbf Q_r +(1-\alpha)\mathbf Q_c -\lambda_{max}(\mathbf J) \mathbf z^{(t)}(\mathbf z^{(t)})^H\right)\mathbf z \right\} \\
&=\text{Re}\left\{\mathbf z^H \left( \mathbf Q -\lambda_{max}(\mathbf J) \mathbf z^{(t)}(\mathbf z^{(t)})^H\right)\mathbf z
\right\}
\end{split}\label{eq:bigproof}
\end{eqnarray}
\hrulefill
\vspace*{0pt}
 \end{figure*}
where
\begin{eqnarray}
\begin{split}
 &\mathbf Q_r = \sum_{k=-L+1}^{L-1} w_k r_{-k}^{(l)} \mathbf A_k =\begin{bmatrix} \sum_{k=-L+1}^{L-1}  w_k r_{-k}^{(l)}\mathbf U_k&\mathbf O\\  \mathbf O& \sum_{k=-L+1}^{L-1}  w_k r_{-k}^{(l)}\mathbf U_k \end{bmatrix},
 \end{split}
\end{eqnarray}
\begin{eqnarray}
\begin{split}
 &\mathbf Q_c = \sum_{k=-L+1}^{L-1} \tilde{w}_k c_{-k}^{(l)} \mathbf B_k
 &=\begin{bmatrix} \mathbf O&   \sum_{k=-L+1}^{L-1}  w_k c_{-k}^{(l)}\mathbf U_k\\  \mathbf O& \mathbf O \end{bmatrix}
 \end{split}
\end{eqnarray}
 \begin{eqnarray}
 \mathbf Q = \alpha\mathbf Q_r + (1-\alpha) \mathbf Q_c.
 \end{eqnarray}

\end{IEEEproof}

\section*{Appendix E\\Proof of theorem 2}
\begin{IEEEproof}
It is well known that FFT(IFFT) computed autocorrelation function is given by \cite{song2016sequence}
\begin{equation}
\begin{split}
 &[C_x^{(t)}(0),C_x^{(t)}(1),\cdots,C_x^{(t)}(L-1),0,C_x^{(t)}(1-L),\cdots,C_x^{(t)}(-1)]^T= \mathbf F^H\!|\mathbf F[(\mathbf y^{(t)})^T\!,\mathbf 0_{1\times L}]^T|^2.
\end{split}
\end{equation}
Similarly,
\begin{equation}
\begin{split}
 &[C_y^{(t)}(0),C_y^{(t)}(1),\cdots,C_y^{(t)}(L-1),0,C_y^{(t)}(1-L),\cdots,C_y^{(t)}(-1)]^T= \mathbf F^H\!|\mathbf F[(\mathbf y^{(t)})^T\!,\mathbf 0_{1\times L}]^T|^2
\end{split}
\end{equation}

Also, $r_{k}^{(t)}=C_x^{(t)}(k)+C_y^{(t)}(k)$ holds, then the following equation is obviously obtained.
\begin{eqnarray}
\begin{split}
&\mathbf r = [r_0^{(t)},r_1^{(t)},\cdots,r_{L-1}^{(t)},0,r_{1-L}^{(t)},\cdots,r_{-1}^{(t)},]\\
&= \mathbf F^H\!|\mathbf F[(\mathbf x^{(t)})^T\!,\mathbf 0_{1\times L}]^T|^2\!+\mathbf F^H\!|\mathbf F[(\mathbf y^{(t)})^T\!,\mathbf 0_{1\times L}]^T|^2.
\end{split}
\end{eqnarray}

The discrete Fourier transform of $x$ and $y$ can be written as
\begin{eqnarray}
	\sum_{l=1}^L x(l)e^{-j2\pi \omega_i l/L} \,\,\mathrm{and}\,\,
\sum_{m=1}^L y(l)e^{-j2\pi \omega_i m/L},
\end{eqnarray}
 respectively. It is easy to verify the following equation holds
\begin{eqnarray}
\begin{split}
&(\sum_{l=1}^L x(l)e^{-j2\pi \omega_i l/L} )^*(\sum_{m=1}^L y(l)e^{-j2\pi \omega_i m/L})=\sum_{l=1}^L \sum_{m=1}^L x^*(l) y(m) e^{-j2\pi \omega_i m/L}\\
&=\sum_{k=-L+1}^{L-1} \sum_{l=0}^{L-1} x^*(l)y(k+l)e^{-j2\pi \omega_i k/L}=\sum_{k=-L+1}^{L-1} c_k e^{-j2\pi \omega_i k/L}
\end{split}
\end{eqnarray}
where $k$ in second equation is to let $k=m-l$.
Then it is easy to know the cross correlation $c_k=C_{xy}(k)$ can be obtained by inverse Fourier transform.

\end{IEEEproof}

\section*{Appendix F\\Proof of Proposition 5}
\begin{IEEEproof}
The objective of (\ref{opt:p04}) can be majorized by
\begin{eqnarray}
\begin{split}
&M_{f_2} (\mathbf z, \mathbf z^{(t)})=\lambda_u \mathbf z^H \mathbf z+2 \text{Re}(\mathbf z^H(\hat{\mathbf Q}-\lambda_u\mathbf I)\mathbf z^{(t)})+(\mathbf z^{(t)})^H(\lambda_u\mathbf I-\mathbf J)\mathbf z^{(t)}
\end{split}
\end{eqnarray}
where $\hat{\mathbf Q} =  \mathbf Q +\mathbf Q^H-2\lambda_{\mathbf J} \mathbf z^{(t)}(\mathbf z^{(t)})^H $. Similarly,

Since $\lambda_u$ should be larger than $\lambda_{max} (\hat{\mathbf Q})$, and $\lambda_{max}(\mathbf Q +\mathbf Q^H)$ is larger than $\lambda_{max} (\hat{\mathbf Q})$, then we can let $\lambda_u > \lambda_{max}(\mathbf Q +\mathbf Q^H)$ so that $\lambda_u>\lambda_{max} (\hat{\mathbf Q})$.

Based on the fact in \cite{song2015optimization}, the following inequality holds
\begin{eqnarray}
\lambda_{max}(\mathbf Q+\mathbf Q^H)\leq 2L\max|a_{ij}|, \quad i,j\in\{1,2,\cdots,2L\}
\end{eqnarray}
where $a_{ij}$ is the elements of the matrix $\mathbf Q+\mathbf Q^H$. Besides, the following inequality holds
$$2\max|Q_{ij}|\geq   \max|Q_{ij}|+\max|Q_{ji}^*|\geq \max|a_{ij}|,$$ and we have $4L (\max \limits_{1\le i,j\le L}| Q_{i,j}|)>2L \max|a_{ij}|$, where $Q_{ij}$ is the elements of $\mathbf Q$.

Hence,  $\lambda_u = 4L (\max \limits_{1\le i,j\le L}|\mathbf Q_{i,j}|)$
\end{IEEEproof}

\section*{Appendix G\\
for calculating  $x = \mathrm{proj}_0(\cdot)$}

Let $$x_k^{(t+1)}=|x_k^{(t+1)}|e^{j \,\mathrm{arg}(\mathbf P_x(\mathbf z^{(t)}))}$$

\begin{itemize}
\item if $m p_c^2 \leq p_e$,
$$\left|x_{k}^{(t+1)}\right|=\left\{\begin{array}{ll}{p_c} & {k=1, \cdots, m} \\ {\sqrt{\frac{p_e-m p_c^{2}}{p_e-m}}} & {k=m+1, \cdots, L}\end{array}\right.$$
\item else if $m p_c^2 > p_e$

$$ \left|x_{k}^{(t+1)}\right|=\min \left\{\delta\left|v_{k}^{(t)}\right|, \gamma\right\}$$
where $\delta \in\left[0, \frac{\gamma}{\min \left\{\left|v_{k}^{(t)}\right|,\left|v_{k}^{(t)}\right| \neq 0\right\}}\right]$ can be obtained by solving the following equation by bisection method.
\begin{eqnarray}
\sum_{n=1}^{N} \min \left\{\delta^{2}\left|v_{k}^{(t)}\right|^{2}, \gamma^{2}\right\}=p_e.
\end{eqnarray}

\end{itemize}
%
\renewcommand{\refname}{References}
\bibliography{myref}
\bibliographystyle{myIEEEtran}

\end{document}